\newif\ifShowKeys
\ShowKeystrue
\ShowKeysfalse

\newif\ifshowtikz
\showtikztrue
\showtikzfalse   

\documentclass[11pt]{article}
	\pdfoutput=1
\usepackage[T1]{fontenc}

\usepackage{listings}
\lstnewenvironment{arkady}  
{\lstset{language=C,frame=trbl,basicstyle = \footnotesize \ttfamily , breaklines = true,showstringspaces=false}}{}

\usepackage{froufrou}
\usepackage{datetime}
\usepackage{comment}					

\usepackage[usenames,dvipsnames]{xcolor}
\definecolor{Maroon}{rgb}{0.15,0.33,0.52}
\definecolor{Mahogany}{rgb}{0.65,0.0975,0.0845}

\usepackage[parsep]{collref}				

\ifShowKeys \usepackage[notcite]{showkeys} \fi

\usepackage{amsmath, amssymb,amsthm}
\usepackage{stackrel}
\numberwithin{equation}{section}
\usepackage{bm,environ,mathrsfs,array,arydshln}
\usepackage{booktabs,float,slashed}
\usepackage{appendix}
\usepackage[mathcal]{euscript}
\usepackage{tensor} 						
\usepackage{mathabx}
\usepackage[vcentermath]{youngtab}
\usepackage{simpler-wick}

\usepackage{graphicx,epsfig,epic}
\usepackage{subcaption,wrapfig}
\usepackage{tikz}
\usepackage{tikz-feynman} 

\allowdisplaybreaks

\usepackage{framed}						
\definecolor{shadecolor}{rgb}{0.9996078, 0.984314, 0.960784}
\definecolor{framecolor}{rgb}{0,0,0}
\definecolor{TFTitleColor}{RGB}{1,1,1}
\definecolor{TFFrameColor}{RGB}{249	218	181}		
\definecolor{TFFrameColor}{RGB}{230 230 230 }

\newenvironment{frshaded}{%
    \MakeFramed {\FrameRestore}}%
    {\endMakeFramed}

\definecolor{myred}{RGB}{233, 33, 45}

\makeatletter
\DeclareFontFamily{OMX}{MnSymbolE}{}
\DeclareSymbolFont{MnLargeSymbols}{OMX}{MnSymbolE}{m}{n}
\SetSymbolFont{MnLargeSymbols}{bold}{OMX}{MnSymbolE}{b}{n}
\DeclareFontShape{OMX}{MnSymbolE}{m}{n}{
    <-6>  MnSymbolE5
   <6-7>  MnSymbolE6
   <7-8>  MnSymbolE7
   <8-9>  MnSymbolE8
   <9-10> MnSymbolE9
  <10-12> MnSymbolE10
  <12->   MnSymbolE12
}{}
\DeclareFontShape{OMX}{MnSymbolE}{b}{n}{
    <-6>  MnSymbolE-Bold5
   <6-7>  MnSymbolE-Bold6
   <7-8>  MnSymbolE-Bold7
   <8-9>  MnSymbolE-Bold8
   <9-10> MnSymbolE-Bold9
  <10-12> MnSymbolE-Bold10
  <12->   MnSymbolE-Bold12
}{}

\let\llangle\@undefined
\let\rrangle\@undefined
\DeclareMathDelimiter{\llangle}{\mathopen}%
                     {MnLargeSymbols}{'164}{MnLargeSymbols}{'164}
\DeclareMathDelimiter{\rrangle}{\mathclose}%
                     {MnLargeSymbols}{'171}{MnLargeSymbols}{'171}
\makeatother

\usepackage{mdframed}

\definecolor{lightpeach}{RGB}{255, 247, 235}

\newmdenv[
  backgroundcolor=gray!5,
  linecolor=gray!40,
  roundcorner=5pt,
  innerleftmargin=8pt,
  innerrightmargin=8pt,
  innertopmargin=6pt,
  innerbottommargin=6pt,
]{remarkbox}

\usepackage{jheppub}
\hypersetup{colorlinks=true,linkcolor=Maroon,citecolor=Maroon,urlcolor=Maroon}

\newcommand{\bs}{\begin{frshaded}}			
\newcommand{\es}{\end{frshaded}\noindent}

\def\ba#1\ea{\begin{align}#1\end{align}}		        
\newcommand{\be}{\begin{equation}}
\newcommand{\ee}{\end{equation}}
\newcommand{\bea}{\begin{equation} \begin{aligned}} 
\newcommand{\eea}{\end{aligned} \end{equation}}
\newcommand{\mc}{\mathcal }
\newcommand{\wh}{\widehat}
\newcommand{\wt}{\widetilde}

\newcommand{\la}{\label}

\newcommand{\lp}{\notag \\ & }

\DeclareMathOperator{\Tr}{\text{Tr}}

\DeclareMathOperator{\tr}{\text{tr}}

\newcommand{\cf}{\textit{cf.} }
\newcommand{\ie}{\textit{i.e.} }
\newcommand{\eg}{\textit{e.g.} }

\renewcommand{\l}{\lambda}

\newcommand{\bJ}{{\mathcal J}}
\newcommand{\sssa}{{\mathsf a}}
\newcommand{\sssb}{{\mathsf b}}
\newcommand{\sssc}{{\mathsf c}}

\begin{document}

\title{The ETH matrix model for DSSYK: non-perturbative corrections and intersection theory}

\author[a,b]{E. Alfinito,}
\author[a,b]{M. Beccaria}

\affiliation[a]{Universit\`a del Salento, Dipartimento di Matematica e Fisica \textit{Ennio De Giorgi}, Lecce, Italy}
\affiliation[b]{INFN - sezione di Lecce, Via Arnesano, I-73100 Lecce, Italy}

\emailAdd{eleonora.alfinito@unisalento.it}
\emailAdd{matteo.beccaria@le.infn.it}

\abstract{
At leading order in the genus expansion the ETH matrix model for DSSYK reproduces its correlators by construction, while its
higher-genus corrections are conjectured to capture higher-topology contributions in the dual sine-dilaton gravity -- 
a correspondence established so far only for the disk and the wormhole. 
At fixed genus the correlators are built from discrete volumes $N_{g,n}$, polynomial in $q$-deformed zeta values 
$\zeta_q(2k)$ with $q=e^{-\lambda}$, $\lambda$ being the DSSYK coupling. These lie in the ring of quasimodular forms generated by the Eisenstein 
series $E_2,E_4,E_6$, whose $S$-duality yields an exact closed form for the leading non-perturbative correction as 
$\lambda\to0$, controlled by $\widetilde q=e^{-4\pi^2/\lambda}$. Known at disk level, this scale is shown here to govern the 
fixed-genus, higher-boundary amplitudes as well. 
We show that the term linear in $\widetilde q$, at leading order in $\lambda$, is captured entirely by the $q$-deformed 
Weil--Petersson volumes, and reduces to a finite sum of intersection 
numbers of $\kappa$-classes on the moduli space $\overline{\mathcal M}_{g,n}$ of stable curves, computable without 
repeating the topological recursion that produced the $N_{g,n}$. We tabulate it for every $(g,n)$ whose $q$-deformed 
volume is known in closed form, and extend it to $(3,1),(3,2),(4,1)$, where none is available. The construction is not 
restricted to leading order: we work out $O(\widetilde q^2)$ for the same cases. 
}


\maketitle


\setcounter{footnote}{0}

\section{Introduction}
\la{sec:intro}

The Sachdev--Ye--Kitaev (SYK) model has been extensively studied as a solvable model of quantum chaos and near-extremal holography 
\cite{Sachdev:1992fk,Maldacena:2016hyu,Maldacena:2016upp,Polchinski:2016xgd}, saturating the chaos bound \cite{Maldacena:2015waa} 
and admitting a genus expansion in $1/N$ analogous to that of random matrix models \cite{Cotler:2016fpe}. 

In the double-scaled limit (DSSYK), the model becomes exactly 
solvable at all energy scales through a chord-diagram expansion \cite{Berkooz:2018jqr,Lin:2022rbf}. 
At the level of the disk, a precise holographic dual has been identified in \cite{Blommaert:2024ymv}: 
sine-dilaton gravity, a two-dimensional dilaton gravity theory with a periodic dilaton potential, whose canonical 
quantization reproduces the $q$-Schwarzian quantum mechanics arising from the chord diagrams of DSSYK, 
with the chord number realized as the (Weyl-rescaled) geodesic length in the bulk.
The periodicity of the potential has a sharp kinematical consequence, worked out in \cite{Blommaert:2024whf}: 
it implies a symmetry under discrete shifts of the momentum conjugate to the length of geodesic slices, and gauging 
this symmetry discretizes the length of the Einstein-Rosen bridge.
Ref.~\cite{Blommaert:2025avl} computed the sine-dilaton 
wormhole (double-trumpet) amplitude via Wheeler--DeWitt quantization
and found that it reproduces exactly the spectral correlation of a finite-cut random matrix integral 
-- identified there with the $q$-deformed JT gravity matrix integral of \cite{Jafferis:2022uhu,Jafferis:2022wez}, 
\ie the ETH matrix model introduced below. In \cite{Blommaert:2025avl}, this is read
as evidence that an all-genus definition of sine-dilaton gravity ought to exist 
and to match the same finite-cut matrix model. However, beyond the disk and wormhole one can no 
longer rely on canonical quantization, and a consistent path-integral prescription would be required.
Whether the identification of higher sine-dilaton gravity correlators with higher-genus (and higher-boundary) matrix model 
amplitudes persists is therefore an open question.

In more detail, let us consider the standard (Majorana) SYK model of $N$ Majorana fermions $\psi_i$, $\{\psi_i,\psi_j\}=2\delta_{ij}$, with Hamiltonian
\be
\la{1.1}
H_{\rm SYK} = i^{p/2}\!\!\sum_{1\leq i_1<\cdots<i_p\leq N}\!\! J_{i_1\cdots i_p}\,\psi_{i_1}\cdots\psi_{i_p},
\ee
where the couplings $J_{i_1\cdots i_p}$ are independent Gaussian random variables of zero mean and variance
\be
\la{1.2}
\llangle J_{i_1\cdots i_p}^2\rrangle_J = \binom Np^{-1},
\ee
following the normalization of \cite{Okuyama:2023kdo}, consistent with (\ref{1.4}) below. 
\footnote{This differs from the more common convention with an explicit $N,p$-independent energy scale $\mathcal J$ held fixed as 
$N,p\to\infty$ -- $\llangle J_{i_1\cdots i_p}^2\rrangle_J=\binom Np^{-1}\frac N{2p^2}\mathcal J^2$ \cite{Sachdev:1992fk,Maldacena:2016hyu,Berkooz:2018jqr} -- 
by a rescaling $J=\mathcal J/\sqrt\lambda$ of the couplings absorbing the double-scaling factor $N/(2p^2)=1/\lambda$; equivalently, (\ref{1.2}) sets $\mathcal J=1$ throughout.}
Let us write $\Tr$
for the unnormalized trace, with $\Tr 1=\dim \mc H_{\rm SYK} = 2^{N/2}$, and $\tr$ for the normalized trace, with $\tr 1 = 1$. 
The double-scaled SYK (DSSYK) model \cite{Berkooz:2018jqr,Lin:2022rbf} is obtained in the limit $N,p\to\infty$ with
\be
\la{1.3}
\lambda\equiv\frac{2p^2}N\ \ \text{ fixed}, \qquad q\equiv e^{-\lambda},
\ee
in which correlators of $H_{\rm SYK}$ reduce to a sum over chord diagrams and become exactly solvable at all energies. In this limit, the DSSYK partition function
reads
\ba
\la{1.4}
& Z_{\rm DSSYK}(\beta) = \lim_{\rm double-scaling}\llangle \tr e^{-\beta H}\rrangle = \int_{0}^{\pi}\frac{d\theta}{2\pi}\mu(\theta)\, e^{-\beta E(\theta)}, \\
\la{1.5}
& E(\theta) = -a\cos\theta, \ \ a = \frac{2}{\sqrt{1-q}},
\ea
where the measure is $\mu(\theta) = (q, e^{\pm 2i\theta}; q)_{\infty}=(q;q)_{\infty}|(e^{2i\theta}; q)|^{2}$.

Refs.~\cite{Jafferis:2022uhu,Jafferis:2022wez} encode these DSSYK thermal correlators in a matrix model based on the Eigenstate Thermalization Hypothesis (ETH).
In this framework, one treats
matrix elements of simple operators between high-energy eigenstates of a chaotic, thermalizing system as pseudo-random variables constrained only 
by coarse data (the density of states and low-point thermal correlators), and since SYK is itself maximally chaotic and thermalizing 
\cite{Maldacena:2015waa,Cotler:2016fpe}, this construction yields an $N_{\rm ETH}\times N_{\rm ETH}$ Hermitian one-matrix model that reproduces (\ref{1.4}) 
at leading order in large $N_{\rm ETH}$. We define
\be
\la{1.6}
\mc Z = \int dM e^{-N_{\rm ETH}\Tr V_{\rm ETH}(M)},\qquad \langle\mc O(M)\rangle = \frac{1}{\mc Z}\int dMe^{-N_{\rm ETH}\Tr V_{\rm ETH}(M)}\,\mc O(M),
\ee
where $\Tr 1 = N_{\rm ETH}$ and $V_{\rm ETH}$ is a $q$-dependent potential fixed by the DSSYK density of states. 
For brevity we refer to this matrix model throughout simply as the ETH matrix model, always meaning this DSSYK-tuned 
case. The genus expansion of correlators in the ETH matrix model reads
\ba
\la{1.7}
\langle \prod_{i=1}^{n}\Tr e^{\beta_{i} M}\rangle_{\rm conn} &= \sum_{g=0}^{\infty}N_{\rm ETH}^{2-2g-n}Z_{g,n}(\beta_{1}, \cdots, \beta_{n}).
\ea
The normalized one-point function $\frac{1}{N_{\rm ETH}}\langle \Tr e^{\beta M}\rangle$ starts with $Z_{0,1}(\beta)$ at large $N_{\rm ETH}$ and by construction
$Z_{0,1}(\beta) = Z_{\rm DSSYK}(\beta)$. A particularly relevant question is whether higher-genus (and higher-boundary) functions $Z_{g,n}(\beta_{1}, \dots, \beta_{n})$ capture
higher correlators in sine-dilaton gravity as suggested in \cite{Blommaert:2025avl}, which verified this assumption for the wormhole, $(g,n)=(0,2)$, by 
Wheeler--DeWitt quantization of the sine-dilaton path integral, as mentioned above.

Regardless of this open gravitational question, the fixed-$(g,n)$ amplitudes
$Z_{g,n}$ define a precise mathematical object whose non-perturbative
dependence on $\l$ can be studied directly. In particular, the
topological recursion representation of these amplitudes involves discrete
volumes $N_{g,n}(b)\equiv N_{g,n}(b_1,\dots,b_n)$ \cite{Okuyama:2023kdo}.
This structure poses a sharp question: what is the leading non-perturbative
correction to a fixed-genus amplitude $Z_{g,n}$ in the semiclassical limit
$\l\to0$?

We find that this correction is controlled by
$\wt q\equiv e^{-4\pi^2/\l}$, and that its coefficient can be
reduced, for arbitrary $(g,n)$, to ordinary $\kappa$-class intersection
numbers on the moduli space of curves. The key point is that, at leading
order as $\l\to0$, this coefficient can be extracted from moduli-space
intersection theory without knowing the full discrete volume.

The argument has three ingredients. First, the quasimodular structure of
the discrete volumes and its modular $S$-transformation determine the
non-perturbative scale. The discrete volumes $N_{g,n}(b)$ are quasi-polynomials
in the squared discrete boundary lengths $b_i^2$, whose coefficients are
polynomials in the $q$-zeta values $\zeta_q(2k)$. These are generating
functions of divisor sums, and as such belong to the ring
of quasimodular forms $\mathbb{Q}[E_2,E_4,E_6]$
\cite{AndrewsRose:2013,BachmannKuhn:2016}; the explicit dictionary is
developed in Section~\ref{sec:quasi-modular}. The $S$-duality transformation
$\tau\to-1/\tau$ then gives a closed-form expression for their leading
non-perturbative correction. The scale
$\wt q=e^{-4\pi^2/\l}$ is already known at disk level, both
from the modular transform of the theta-function measure
\cite{Okuyama:2025fhi} and from the $E_2,E_4,E_6$ ring itself
\cite{Beccaria:2026ndg}; what is new here is that the same transformation
acts on fixed-genus discrete volumes and can be evaluated in closed form.
Because it acts the same way on every $\zeta_q(2k)$, the resulting shift is
universal: the topology enters only through how it acts on the volume.

Second, the degree hierarchy isolates the relevant part of the discrete
volume: the leading correction is captured entirely by the $q$-deformed
Weil--Petersson volumes $V^q_{g,n}(L_1,\dots,L_n)$ of \cite{DoNorbury:2025qmirz,DoNorbury:2025dssyk}, functions of 
continuous boundary lengths $L_i$. Third, the
intersection-theoretic representation of $V^q_{g,n}(L)$ converts the
universal modular shift into a finite sum of ordinary $\kappa$-class
intersection numbers, computable without repeating the topological
recursion that produced $N_{g,n}(b)$.

The quasi-polynomial and $q\to1$ properties underlying this construction
have been established independently by
\cite{DoNorbury:2025dssyk,Giacchetto:2025nnb}; our analysis takes these
results as input and studies their non-perturbative dependence
on $\l$.
Crucially, the method does not require prior knowledge of the full
$q$-deformed volume. We therefore apply it to three genuinely new cases,
$(g,n)=(3,1),(3,2),(4,1)$,  
beyond the cases for which it is currently known in closed form,
and obtain
their leading non-perturbative corrections directly from intersection
theory. Finally, at $O(\wt q^2)$ a second universal shift appears,
together with mixed two-insertion $\kappa$-class intersection numbers,
providing the first step toward a systematic higher-order non-perturbative
expansion at fixed genus.

\paragraph{Plan of the paper} Sections~\ref{sec:ETH-genus} and \ref{sec:quasi-modular} recall the ETH matrix model for DSSYK 
and discrete volumes, and develop the quasimodular structure of the $\zeta_q(2k)$. 
Section~\ref{sec:N11-N12} illustrates the leading non-perturbative correction directly at $(1,1)$ and $(1,2)$, and 
Section~\ref{sec:bridge} shows by degree counting that it is governed by the $q$-deformed Weil--Petersson volumes. 
Sections~\ref{sec:constant} and \ref{sec:nonzero} derive the general closed form, at $L=0$ and for the full 
$L$-dependence; Section~\ref{sec:beyond-DN} applies it to $(3,1)$, $(3,2)$ and $(4,1)$, and Section~\ref{sec:NP2} 
carries the construction to $O(\wt q^2)$. Sections~\ref{sec:comments} and \ref{sec:summary} compare with related 
non-perturbative results and summarise. Four appendices give an independent check at $(2,1)$ 
(Appendix~\ref{app:N21}), the proofs of the closed-form shift operators (Appendix~\ref{app:deltas}), the algorithm for 
the $\kappa$-class intersection numbers (Appendix~\ref{app:DVV}), and the constant terms at $(0,6)$ and $(2,2)$ 
(Appendix~\ref{app:discrepancy}).

\section{Structure of the ETH matrix model genus expansion}
\la{sec:ETH-genus}

Eynard--Orantin topological recursion \cite{Eynard:2007kz}, applied in \cite{Okuyama:2023kdo} to the spectral data of the ETH matrix model, 
yields the representation
\be
\la{2.1}
Z_{g,n}(\beta_{1}, \cdots, \beta_{n}) = \sum_{b_{1}, \dots, b_{n}\in\mathbb Z_{+}}N_{g,n}(b_{1}, \cdots, b_{n})\prod_{i=1}^{n}b_{i}\, I_{b_{i}}(a\beta_{i}),
\ee
where $I_{b}$ are modified Bessel functions and the parameter $a$ was introduced in (\ref{1.5}). The first few cases 
of $N_{g,n}(b_{1}, \dots, b_{n})$ are computed explicitly in \cite{Okuyama:2023kdo}. Let us introduce the rescaled quantity
\be
\la{2.2}
\wt N_{g,n} = (q;q)_{\infty}^{3(2g-2+n)}N_{g,n},
\ee
where 
\be
\la{2.3}
(q;q)_{\infty} = \prod_{n=1}^{\infty}(1-q^{n}).
\ee
This rescaling has a curve-level origin.\footnote{Recall that Eynard--Orantin topological recursion builds, from a spectral curve $(x(z),y(z))$ on $\mathbb{CP}^1$ together with the 
Bergman kernel $B(z_1,z_2)=dz_1dz_2/(z_1-z_2)^2$, a tower of multi-differentials $\omega_{g,n}(z_1,\dots,z_n)$ for every $(g,n)$ with $2g-2+n>0$, 
defined recursively by a residue at the branch points $z=\alpha$ (the zeros of $dx$) of a kernel $K(z_0,z)\,\propto\,1/y(z)$; the discrete volumes $N_{g,n}(b_1,\dots,b_n)$ 
are then read off as the coefficients of $\omega_{g,n}$ in the basis $\prod_ib_iz_i^{b_i-1}dz_i$, cf.\ (\ref{2.1}). The spectral curve considered in \cite{Okuyama:2023kdo} 
is $x(z)=\frac a2(z+z^{-1})$, $y(z)=\frac1a(z-z^{-1})\prod_{k\geq1}(1-q^k)(1-z^2q^k)(1-z^{-2}q^k)$. At the branch points $z=\pm1$, $y(z)$ vanishes linearly, 
with leading coefficient $y'(\pm1)\propto\prod_k(1-q^k)^3=(q;q)_\infty^3$. Each of the $2g-2+n$ recursive steps needed to reach $\omega_{g,n}$ contributes 
one inverse power of this residue, so $N_{g,n}$ carries an overall $(q;q)_\infty^{-3(2g-2+n)}$, and \eqref{2.2} is exactly the rescaling of $\omega_{g,n}$, 
equivalently of $y(z)$ itself, that removes it.} 
Since $(q;q)_\infty\to0$ exponentially as $\l\to0$ (see \eqref{3.11} below), the raw $N_{g,n}$ carries an exponentially 
divergent prefactor. Removing it leaves $\wt N_{g,n}$ with only a power-law divergence $\sim\l^{-(6g-6+2n)}$, matching 
that of $V^q_{g,n}$ -- this is what makes $\wt N_{g,n}$, rather than $N_{g,n}$, the natural object to compare against 
the $V^q_{g,n}$ studied in Section~\ref{sec:bridge} and beyond.
One has, for instance, 
\be
\la{2.4}
\wt N_{1,1} = \bigg[\frac{b^{2}-4}{48}+\frac{\zeta_{q}(2)}{2}\bigg]\,P_{b}, \qquad P_{b}\equiv \frac{1+(-1)^{b}}{2},
\ee
where the $\zeta_{q}(s)$ quantity is 
\be
\la{2.5}
\zeta_{q}(s) = \sum_{n=1}^{\infty}(q^{-\frac{1}{2}n}-q^{\frac{1}{2}n})^{-s}.
\ee
More generally, as observed in \cite{Okuyama:2023kdo} and later proven in \cite{DoNorbury:2025dssyk},  $\wt N_{g,n}(b_1,\dots,b_n)$ is always a 
quasi-polynomial in $b_1^2,\dots,b_n^2$,
\ie when restricted to each parity class of $b_1,\dots,b_n$, it is a (generically different) polynomial.
The coefficients are polynomial in the even $q$-zeta values $\zeta_q(2),\zeta_q(4),\zeta_q(6),\dots$ of (\ref{2.5}). If $b_i$ and $\zeta_q(2k)$ 
are assigned the weighted degrees $\deg b_i=1$, $\deg\zeta_q(2k)=2k$, this quasi-polynomial has degree bounded by $6g-6+2n$, with $b_i^2$ and 
$\zeta_q(2k)$ genuinely mixing at every allowed degree. A more elaborate example than (\ref{2.4}),
illustrating these general features, is 
\ba
\la{2.6}
\wt N_{1,2}(b_{1},b_{2}) &= \bigg[\frac{(\bm{b}^{2}-10)(\bm{b}^{2}-2)}{384}+\frac{\bm{b}^{2}-2}{4}\zeta_{q}(2)+\frac{7}{2}\zeta_{q}(2)^{2}+\frac{5}{2}
\zeta_{q}(4)\bigg]P_{b_{1}+b_{2}}+\frac{1}{32}P_{b_{1}}P_{b_{2}},
\ea
where $\bm{b}^{2}=b_{1}^{2}+b_{2}^{2}$.

\section{Quasimodular properties of $\zeta_{q}(2n)$ and their $q\to 1$ expansion}
\la{sec:quasi-modular}

The coefficients of the discrete volumes \eqref{2.4}, \eqref{2.6} are polynomials in the $q$-zeta values 
$\zeta_{q}(2n)$ of \eqref{2.5}, so their behaviour as $\l\to0$ is entirely controlled by the behaviour of the 
$\zeta_{q}(2n)$ themselves. That limit is singular: $q=e^{-\l}\to1$, which is the boundary of the domain of 
convergence of \eqref{2.5}, and the naive expansion of the summand term by term does not converge.

The way around this is that the $\zeta_{q}(2n)$ are not arbitrary functions of $q$. They are generating functions 
of divisor sums, and therefore lie in the ring of quasimodular forms generated by the Eisenstein series 
$E_{2},E_{4},E_{6}$. Setting $q=e^{2\pi i\tau}$, the $\l\to0$ limit is $\tau\to0$, which the modular 
$S$-transformation $\tau\to-1/\tau$ maps to the cusp $\tau\to i\infty$, where the $q$-expansion converges rapidly. 
Applying $S$ therefore reorganizes each $\zeta_{q}(2n)$ into a finite sum of terms with explicit powers of $1/\l$ 
multiplying Eisenstein series evaluated at $\wt q = e^{-4\pi^{2}/\l}$,
the image of $q$ under the transformation. Since $E_{2n}(\wt q)=1+O(\wt q)$, this separates the perturbative 
small-$\l$ tower from the non-perturbative corrections, and identifies $\wt q$ as the scale controlling the latter.

\subsection{Eisenstein series}

The Eisenstein series are defined by the following expansion for $|q|<1$
\be
\la{3.1}
E_{2m}=1-\frac{4m}{B_{2m}}\sum_{n=1}^{\infty}\frac{n^{2m-1}q^{n}}{1-q^{n}}, \qquad m=1,2,\dots,
\ee
where $B_{2m}$ are the Bernoulli numbers. Equivalently, using the divisor sum $\sigma_k(n)\equiv\sum_{d|n}d^k$ and the standard 
rearrangement $\sum_{n\geq1}\frac{n^kq^n}{1-q^n}=\sum_{n,j\geq1}n^kq^{nj}=\sum_{N\geq1}q^N\sum_{n|N}n^k=\sum_{N\geq1}\sigma_k(N)q^N$,
we get 
\be
\la{3.2}
E_{2m}=1-\frac{4m}{B_{2m}}\sum_{n=1}^{\infty}\sigma_{2m-1}(n)\,q^{n}.
\ee
The first three cases 
\be
E_{2}=1-24\sum_{n=1}^{\infty}\frac{n\,q^{n}}{1-q^{n}}, \qquad
E_{4}=1+240\sum_{n=1}^{\infty}\frac{n^{3}\,q^{n}}{1-q^{n}}, \qquad
E_{6}=1-504\sum_{n=1}^{\infty}\frac{n^{5}\,q^{n}}{1-q^{n}},
\ee
are particularly important
because any higher $E_{2n}$ can be expressed as a polynomial in $E_{4}$ and $E_{6}$, as follows from the structure theorem for modular forms of $SL(2,\mathbb Z)$. The first examples are 
\be
\la{3.4}
E_{8}=E_{4}^{2},\qquad E_{10}=E_{4}E_{6},\qquad E_{12}=\frac{441\,E_{4}^{3}+250\,E_{6}^{2}}{691}, \qquad E_{14}=E_{4}^{2}E_{6}.
\ee
Under the $SL(2, \mathbb Z)$ transformation
\be
\tau\to \tau' = \frac{a\tau+b}{c\tau+d}, \qquad \begin{pmatrix} a & b \\ c & d \end{pmatrix}\in SL(2,\mathbb Z),
\ee
the Eisenstein series transform according to 
\bea
\la{3.6}
E_{2}(\tau') &= (c\tau+d)^{2}E_{2}(\tau)+\frac{6}{\pi i}c(c\tau+d), \\
E_{4}(\tau') &= (c\tau+d)^{4}E_{4}(\tau), \qquad
E_{6}(\tau') = (c\tau+d)^{6}E_{6}(\tau),
\eea
where the notation is $E_{n}(\tau)\equiv E_{n}(e^{2\pi i \tau})$. The relations  (\ref{3.6}) show that $E_{2}$ is quasimodular, while $E_{4}$ and $E_{6}$
are genuine modular forms. In particular, for the inversion transformation $\tau\to -1/\tau$, one has
\ba
\la{3.7}
E_{2}(-1/\tau) &= \tau^{2}E_{2}(\tau)-\frac{6i\tau}{\pi}, \qquad
E_{4}(-1/\tau) = \tau^{4}E_{4}(\tau), \qquad
E_{6}(-1/\tau) = \tau^{6}E_{6}(\tau).
\ea
The closely related $q$-function $(q;q)_{\infty}$ in (\ref{2.3}) may be expressed in terms of the Dedekind $\eta$ function 
\be
(q;q)_{\infty} = q^{-1/24}\eta(\tau), 
\ee
with modular inversion transformation rule
\be
\la{3.9}
\eta(-1/\tau) = \sqrt{-i\tau}\, \eta(\tau).
\ee
In our context $q=e^{-\l} = e^{2\pi i \tau}$, \ie $\tau=\frac{i}{2\pi}\l$. Thus $\tau' = -1/\tau = \frac{2\pi i}{\l}$ and we get the associated parameter
\be
\la{3.10}
\wt q = e^{2\pi i \tau'} = e^{-\frac{4\pi^{2}}{\l}}.
\ee
The inversion formulas (\ref{3.7}) and (\ref{3.9}) then give 
\bea
\la{3.11}
E_{2}(q) &= -\bigg(\frac{2\pi}{\l}\bigg)^{2}E_{2}(\wt q)+\frac{12}\l, \qquad
E_{4}(q) = \bigg(\frac{2\pi}{\l}\bigg)^{4}\,E_{4}(\wt q), \qquad
E_{6}(q) =-\bigg(\frac{2\pi}{\l}\bigg)^{6}\,E_{6}(\wt q),  \\ 
(q;q)_{\infty} &= \sqrt{\frac{2\pi}\l}\,e^{-\pi^{2}/(6\l)+\l/24}\,(\wt q;\wt q)_{\infty}.
\eea
The expansion of these expressions in powers of $\wt q$ provides all non-perturbative corrections in the small-$\l$ limit. 

\subsection{Expressing $\zeta_{q}(2n)$ in terms of Eisenstein series}

All $\zeta_{q}(2n)$ can be expressed in terms of Eisenstein series. This is an instance of a standard phenomenon: $q$-analogues of zeta 
values built from Lambert-type sums are generating functions of divisor sums, and the algebra they generate is the ring of quasimodular 
forms \cite{AndrewsRose:2013,BachmannKuhn:2016}. 

In more detail, let us begin with the case of $\zeta_{q}(2)$. Writing $\zeta_{q}(2)=\sum_{m\geq1}\frac{q^{m}}{(1-q^{m})^{2}}$
and expanding $\frac{x}{(1-x)^{2}}=\sum_{k\geq1}kx^{k}$ gives the classical Lambert-series identity
\be
\la{3.12}
\zeta_{q}(2) = \sum_{m,k\geq1}k\,q^{mk} = \sum_{N=1}^{\infty}\sigma_{1}(N)\,q^{N} = \frac{1-E_{2}}{24},
\ee
where we used (\ref{3.2}). The next case is $\zeta_{q}(4)$. The same Lambert-series method, applied to
$\frac{x^{2}}{(1-x)^{4}}=\sum_{j\geq1}\frac{j^{3}-j}6x^{j}$, gives
\be
\la{3.13}
\zeta_{q}(4) = \sum_{m\geq1}\frac{q^{2m}}{(1-q^{m})^{4}} = \frac16\Big[\sum_{N=1}^{\infty}\sigma_{3}(N)q^{N}-\sum_{N=1}^{\infty}\sigma_{1}(N)q^{N}\Big]
= \frac{E_{4}+10E_{2}-11}{1440}.
\ee
Extension to any $\zeta_{q}(2n)$ is similar. Indeed, we have
\be
\la{3.14}
\zeta_{q}(2n) = \sum_{m\geq1}\frac{q^{nm}}{(1-q^{m})^{2n}} =\sum_{m,j\geq1}\binom{j+n-1}{2n-1}q^{mj}.
\ee
The binomial coefficient is polynomial in $j$, and each power $j^{p}$ appearing in it can be replaced, term by term, using
\be
\la{3.15}
\sum_{m,j\ge 1}j^{p}q^{mj} = \sum_{N=1}^{\infty}\sigma_{p}(N)q^{N},
\ee
and use (\ref{3.2}) to express $\zeta_q(2n)$ in terms of Eisenstein sums. For example,
from $\frac{x^{3}}{(1-x)^{6}}=\sum_{j}\binom{j+2}5x^{j}$ and
$\frac{x^{4}}{(1-x)^{8}}=\sum_{j}\binom{j+3}7x^{j}$, one finds -- recall the ring relations (\ref{3.4}) -- 
\be
\la{3.16}
\zeta_{q}(6) = \frac{191 - 168E_{2} - 21E_{4} - 2E_{6}}{120960}, \qquad
\zeta_{q}(8) = \frac{2160E_{2}+294E_{4}+40E_{6}+3E_{4}^{2}-2497}{7257600}.
\ee
The mechanism is transparent from (\ref{3.14})--(\ref{3.15}): the binomial coefficient is a polynomial in $j$ of degree $2n-1$, so 
$\zeta_q(2n)$ is a rational combination of the divisor generating functions $\sum_N\sigma_p(N)q^N$ with $p\leq 2n-1$, each of which is 
quasimodular of weight $p+1$. Hence $\zeta_q(2n)$ is quasimodular of weight at most $2n$, which is precisely the weight assignment used 
in the degree counting of Section~\ref{sec:bridge}.

Using the inversion relations \eqref{3.11}, we get the following small-$\l$ expansions of $\zeta_{q}(2n)$
\bea
\la{3.17}
\zeta_{q}(2) &= \frac1{24}\Big(\frac{2\pi}\l\Big)^{2}E_{2}(\wt q) - \frac1{2\l}+\frac1{24}, \\
\zeta_{q}(4) &= \frac1{1440}\Big(\frac{2\pi}\l\Big)^{4}E_{4}(\wt q) - \frac1{144}\Big(\frac{2\pi}\l\Big)^{2}E_{2}(\wt q) + \frac1{12\l}-\frac{11}{1440}, \\
\zeta_{q}(6) &= \frac1{60480}\Big(\frac{2\pi}\l\Big)^{6}E_{6}(\wt q) - \frac1{5760}\Big(\frac{2\pi}\l\Big)^{4}E_{4}(\wt q) + \frac1{720}\Big(\frac{2\pi}\l\Big)^{2}E_{2}(\wt q) - \frac1{60\l}+\frac{191}{120960}, \\
\zeta_{q}(8) &= \frac1{2419200}\Big(\frac{2\pi}\l\Big)^{8}E_{4}(\wt q)^{2} + \frac7{172800}\Big(\frac{2\pi}\l\Big)^{4}E_{4}(\wt q) - \frac1{181440}\Big(\frac{2\pi}\l\Big)^{6}E_{6}(\wt q)  \\
& - \frac1{3360}\Big(\frac{2\pi}\l\Big)^{2}E_{2}(\wt q) + \frac1{280\l}-\frac{2497}{7257600}.
\eea
In each case, the terms independent of $\wt q$ form the purely algebraic (perturbative) small-$\l$ tower, while every $E_{2n}(\wt q)$ carries the full non-perturbative content, $E_{2n}(\wt q)=1+O(\wt q)$.

\section{Non-perturbative corrections to $\wt N_{1,1}$ and $\wt N_{1,2}$}
\la{sec:N11-N12}

Substituting \eqref{3.17} into (\ref{2.4}) and \eqref{2.6}, and extracting the term linear in $\wt q$ -- exact in $\l$, 
not just its leading power -- shows directly, from degree-counting alone, that the leading small-$\l$ non-perturbative correction comes 
entirely from the top-degree part of each quasi-polynomial. Let us illustrate this explicitly in the two examples above, for clarity. 

In the case of $\wt N_{1,1}$,
we have $(g,n)=(1,1)$, and the top degree is $6g-6+2n=2$. In (\ref{2.4}), $\frac{b^2}{48}+\frac{\zeta_q(2)}2$ 
has degree $2$ (top) and $-\frac1{12}$ has degree $0$; the latter is $\wt q$-independent and drops out entirely. So
\be
\la{4.1}
\big[\wt N_{1,1}(b)\big]_{O(\wt q)} = \frac12\big[\zeta_q(2)\big]_{O(\wt q)} = -\frac{2\pi^2}{\l^2}, \qquad b\text{ even}.
\ee
This is exact and $b$-independent, and comes entirely from the top-degree piece (the only other piece is $\wt q$-independent).
Moving to the second example, $\wt N_{1,2}$, we have $(g,n)=(1,2)$, and the top degree is $6g-6+2n=4$. Here, the split is less trivial. Writing $\bm b^2=b_1^2+b_2^2$, the bracket of \eqref{2.6} decomposes as
\be
\underbrace{\frac{\bm b^4}{384}+\frac{\bm b^2}4\zeta_q(2)+\frac72\zeta_q(2)^2+\frac52\zeta_q(4)}_{\text{degree }4\text{ (top)}} \ 
+\ \underbrace{-\frac{\bm b^2}{32}-\frac12\zeta_q(2)+\frac5{96}}_{\text{degree}\leq2}.
\ee
Substituting \eqref{3.17} and extracting $O(\wt q)$ from each piece separately gives\footnote{
This is for $b_1,b_2$ both even or both odd; the $\frac1{32}P_{b_1}P_{b_2}$ piece of \eqref{2.6} is $\wt q$-independent throughout.}
\be
\la{4.3}
\big[\text{top}\big]_{O(\wt q)} = \frac{2\pi^4}{\l^4}+\frac{14\pi^2}{\l^3}+\frac{\pi^2}{2\l^2}-\frac{\pi^2}{\l^2}\bm b^2, \qquad
\big[\text{lower}\big]_{O(\wt q)} = \frac{2\pi^2}{\l^2},
\ee
Let us look at the $b$-independent pieces. The leading one at small $\l$ is $2\pi^{4}/\l^{4}$ and comes from the top-degree part. 
The $b$-dependent part has a leading contribution $-\pi^{2}/\l^{2}\bm{b}^{2}$ that also comes from the top-degree part. Both have degree 4 if we formally
assign $\deg\l=-1$ after the small-$\l$ expansion, \ie at the level of (\ref{4.3}). The degree assignment is summarized in 
\be
\la{4.4}
\def\arraystretch{1.3}
\begin{array}{cccccc}
\toprule
& \zeta_{q}(2k) & & b_{i} & & \l \\
\midrule
\text{degree} & 2k && 1 && -1 \\
\bottomrule
\end{array} \ .
\ee
With this assignment the leading small-$\l$ behaviour $\zeta_q(2k)\sim\zeta(2k)\,\l^{-2k}$ is degree-preserving, so the 
weighted degree of Section~\ref{sec:ETH-genus} is unchanged by the expansion.
Collecting the leading terms, we obtain
\be
\la{4.5}
\big[\wt N_{1,1}(b)\big]_{O(\wt q)}^{\rm LO} = -\frac{2\pi^2}{\l^2}, \qquad 
\big[\wt N_{1,2}(b_1,b_2)\big]_{O(\wt q)}^{\rm LO} = \frac{2\pi^4}{\l^4}-\frac{\pi^2}{\l^2}\,\bm b^2,
\ee
where the superscript ``LO'' denotes the leading small-$\l$ term at $O(\wt q)$.

The same analysis can be carried out for $\wt N_{2,1}$, where the structure is considerably richer: the top degree is 
$6g-6+2n=8$, and the quasi-polynomial contains lower-degree components at degrees $6,4,2$ and $0$, rather than the single 
lower component of the $(1,2)$ case. Appendix~\ref{app:N21} shows that each component contributes at exactly $\l^{-d'}$, with 
$d'$ its own degree, so that the four lower components are suppressed by $\l^2,\l^4,\l^6,\l^8$ relative to the top one; the 
resulting leading correction reproduces the $(2,1)$ entries of Tables~\ref{tab:322a} and \ref{tab:47a}.

Reaching \eqref{4.5} already required knowledge of the full quasi-polynomials \eqref{2.4}, \eqref{2.6} -- themselves the outcome of running topological recursion on the ETH matrix model 
spectral curve, a calculation that grows rapidly with $(g,n)$. The goal of the rest of this paper is to bypass that step: rather than running topological recursion 
on the  
$q$-dependent spectral curve of the ETH matrix model, we show that the same combined-degree-$D$ leading correction, for arbitrary $(g,n)$, is computable from 
two decoupled and much simpler ingredients -- the universal (Witten--Kontsevich) intersection numbers of the moduli space $\overline{\mc M}_{g,n}$, 
and a closed-form, purely modular expression for the $q$-dependence. Note that 
the former are themselves generated by a recursion -- the Dijkgraaf--Verlinde--Verlinde/DVV recursion -- but on the trivial Airy curve rather than that of the ETH model, 
and are purely combinatorial, with no $q$-dependence anywhere.

\section{From DSSYK discrete volumes to $V^q_{g,n}(L)$}
\la{sec:bridge}

The classical Weil--Petersson volume $V^{\rm WP}_{g,n}(L_1,\dots,L_n)$ is the symplectic volume of the moduli space $\overline{\mc M}_{g,n}$ of genus-$g$ hyperbolic surfaces 
with $n$ labeled geodesic boundaries of lengths $L_1,\dots,L_n$.
\footnote{A hyperbolic surface is a Riemann surface equipped with a complete metric of constant curvature $-1$. 
By the uniformization theorem, every surface of negative Euler characteristic admits a unique such metric, so ``hyperbolic surface'' and ``Riemann surface'' 
(with the geodesic boundaries playing the role of the marked points, in the $L\to0$ limit) can be used interchangeably here. 
As usual, $\overline{\mc M}_{g,n}$ is the Deligne--Mumford compactification of this moduli space, of complex dimension 
$3g-3+n$. Concretely, one adds the degenerate surfaces obtained by pinching a simple closed geodesic to zero length: 
the pinch either separates the surface into two stable pieces or is non-separating and lowers the genus by one. This 
recursive boundary structure is what underlies the recursions used below, and compactness is what makes the volumes 
finite. For a physicist-oriented introduction to moduli spaces of Riemann surfaces, their boundary structure and the 
associated intersection theory, see \cite{Giacchetto:2024aka} and Appendix A in \cite{do2011}.
} 
Mirzakhani showed that it is a polynomial 
in $L_1^2,\dots,L_n^2$, with coefficients given by intersection numbers on $\overline{\mc M}_{g,n}$ \cite{Mirzakhani:2006fta,Mirzakhani:2006eta}. 
Reference \cite{DoNorbury:2025qmirz} introduces a $q$-deformation of this quantity, $V^{q}_{g,n}(L_{1}, \dots, L_{n})$, 
depending on continuous variables $L_{i}$. This quantity is different from $\wt N_{g,n}$
but related to it by Theorem 1.3 in \cite{DoNorbury:2025dssyk}\footnote{
An independent proof that the ETH matrix integral for DSSYK furnishes a discrete $q$-analogue of the Weil--Petersson 
volumes -- \ie of the $q\to1$ statement, rather than of the top-degree relation \eqref{5.1} -- was given in 
\cite{Giacchetto:2025nnb}, using the Kontsevich--Soibelman quantum Airy structure formulation of topological recursion 
in place of the Chekhov--Eynard--Orantin one. Their analysis applies to pruned correlators of generic one-cut matrix 
models, of which discrete volumes are one example.}
\be
\la{5.1}
\wt N_{g,n}(b_1,\dots,b_n) = 2^{3-2g-n}\,V^{q}_{g,n}(b_1,\dots,b_n) + [\text{terms of strictly lower degree}],
\ee
using the weighted degree $\deg b_i^2=\deg L_i^2=2$, $\deg\zeta_q(2k)=2k$, in which $V^q_{g,n}(L)$ is exactly homogeneous of 
degree $D=6g-6+2n$. In other words, $V^q_{g,n}$ is evaluated by formally substituting the continuous $L_i$ for the integers $b_i$ and keeping 
 (up to the stated normalization) the top-degree part of $\wt N_{g,n}$ -- with genuine, generally nonzero corrections at every strictly lower degree.
\footnote{
This is already directly checkable at $(1,1)$: comparing \eqref{2.4} ($b$ even) to $V^q_{1,1}(L)=\frac1{48}L^2+\frac12\zeta_q(2)$ 
(taken from \cite{DoNorbury:2025dssyk}'s own appendix table; here $2^{3-2g-n}=2^0=1$) gives $\wt N_{1,1}(b) - V^{q}_{1,1}(b) = -\frac1{12}$,
exactly: the top-degree $b^2/48$ terms cancel identically, leaving only a degree-$0$ remainder. }

The fact that $V^{q}_{g,n}$ captures only the top-degree part of $\wt N_{g,n}$ is enough for our purposes: it lets the general machinery 
developed below reproduce, for arbitrary $(g,n)$, what the direct calculation of (\ref{4.5}) gave explicitly for $(1,1)$ and $(1,2)$. Let us begin with the case $b_{i}=0$.
 $V^q_{g,n}(0)$ consists entirely of $\zeta_q$-monomials of top degree $D$; by \eqref{5.1}, the terms in $\wt N_{g,n}(0)$  
 beyond $V^q_{g,n}(0)$ all sit at strictly lower degree, hence scale as $\l^{-(D-2)}$ or weaker as $\l\to0$ -- strictly subleading to 
 $V^q_{g,n}(0)$'s own $\l^{-D}$ divergence, at every order in $\wt q$,
 whenever the top-degree coefficient is itself nonzero -- as it is in every case computed below. 
  So the \emph{leading} $\l$-power of the leading non-perturbative correction
 is exactly the leading $\l$-power of the true $\wt N_{g,n}(0)$ as well.
 The same argument applies term by term to the general $L$-dependence of $V^q_{g,n}(L)$: each $L^{2\alpha}$-coefficient is itself exactly homogeneous of 
 degree $D-2|\alpha|$, so its own leading $\l^{-(D-2|\alpha|)}$ term is likewise protected.

Section~\ref{sec:N11-N12} already shows what this leaves out: subleading-in-$\l$ corrections would need the 
lower-degree remainder's own non-perturbative structure, which is not currently known in general. This is why we will not pursue subleading-$\l$ 
corrections to $V^q_{g,n}(L)$ beyond the leading order established above.

As a final remark, let us comment on the $(q;q)_\infty$ rescaling factor in (\ref{2.2}). From that relation we recover $N_{g,n}$ from $\wt N_{g,n}$ via
\be
\la{5.2}
N_{g,n}(b) = (q;q)_\infty^{-3(2g-2+n)}\,\wt N_{g,n}(b),
\ee
and $(q;q)_\infty$ itself has an exact, independently non-perturbative structure given in the last equation of (\ref{3.11}). Writing  $\nu\equiv3(2g-2+n)$ and using the infinite-product form there,
\be
\la{5.3}
(q;q)_\infty^{-\nu} = P(\l)\big[1+\nu\,\wt q+O(\wt q^2)\big], \qquad P(\l)\equiv\Big(\frac{2\pi}\l\Big)^{-\nu/2}e^{\nu\left[\pi^2/(6\l)-\l/24\right]},
\ee
which is a clean multiplicative correction. Concretely,
\be
\la{5.4}
\big[N_{g,n}(0)\big]_{O(\wt q)} = P(\l)\Big\{\nu\,\big[\wt N_{g,n}(0)\big]_{\rm cl} + \big[\wt N_{g,n}(0)\big]_{O(\wt q)}\Big\}.
\ee
Both terms inside the braces are pinned to the \emph{same} leading power of $\l$, by the same mechanism as before.\footnote{In more detail, 
$[\wt N_{g,n}(0)]_{\rm cl}$ decomposes into pieces of degree $d'\leq D$, each scaling as $\l^{-d'}$, so its leading small-$\l$ term is 
necessarily the top ($d'=D$) piece -- exactly $2^{3-2g-n}V^{WP}_{g,n}(0)$ by Theorem 1.2 in \cite{DoNorbury:2025dssyk}. 
The same is true of $[\wt N_{g,n}(0)]_{O(\wt q)}$: its leading term is likewise the top-degree piece, exactly $2^{3-2g-n}[V^q_{g,n}(0)]_{O(\wt q)}^{\rm LO}$. }
The sum in (\ref{5.4}) then gives
\be
\la{5.5}
\big[N_{g,n}(0)\big]_{O(\wt q)}^{\rm LO} = 2^{3-2g-n}\,P(\l)\,\l^{-D}\Big[3(2g-2+n)\,V^{\rm WP}_{g,n}(0)+C_{g,n}\Big]\wt q+\cdots,
\ee
where $[V^q_{g,n}(0)]_{O(\wt q)}^{\rm LO}\equiv C_{g,n}\,\l^{-D}\wt q$, with $C_{g,n}=\pi^{2d}\sssb_{g,n}$ in the normalization of \eqref{6.19}, is the quantity we will compute later in 
Sections~\ref{sec:constant}--\ref{sec:nonzero}, and $V^{\rm WP}_{g,n}(0)$ is $V^{\rm WP}_{g,n}(L)$ at $L=0$, i.e.\ the (rescaled) $q\to1$ limit of $V^q_{g,n}(0)$.

From this point on, $L_i$ denotes specifically the continuous variable of $V^q_{g,n}(L)$, to match the notation in \cite{DoNorbury:2025dssyk}; it coincides with the discrete $b_i$ 
integers only in the top-degree sense of \eqref{5.1}.

\section{Leading non-perturbative correction to $V^{q}_{g,n}(0)$}
\la{sec:constant}

Following \cite{DoNorbury:2025dssyk}, the $q$-deformed Weil--Petersson volumes are
\be
\la{6.1}
V^{q}_{g,n}(L_{1},\dots,L_{n}) = \int_{\overline{\mc M}_{g,n}}\exp\Big(\Omega_{q}(\kappa_{1},\kappa_{2},\dots)+\frac12\sum_{i=1}^{n}L_{i}^{2}\psi_{i}\Big),
\qquad \Omega_{q}=\sum_{m\geq1}s_{m}(q)\,\kappa_{m}.
\ee
Here $\psi_i$ and $\kappa_m$ are standard cohomology classes on $\overline{\mc M}_{g,n}$, generating its so-called tautological ring: 
$\psi_i$ is associated to the $i$-th marked point, and $\kappa_m$ is a further, closely related class obtained from the $\psi$'s by a standard integration (``pushforward'') procedure.
\footnote{Precisely: $\psi_i\in H^2(\overline{\mc M}_{g,n};\mathbb Q)$ is the first Chern class of the line bundle whose fiber, 
at each point of moduli space, is the cotangent line to the curve at the $i$-th marked point; $\kappa_m\in H^{2m}(\overline{\mc M}_{g,n};\mathbb Q)$ 
is the Mumford--Morita--Miller class $\kappa_m=\pi_*\big(\psi_{n+1}^{m+1}\big)$, obtained by pushing the $(m{+}1)$-th power of $\psi_{n+1}$ 
forward along the map $\pi:\overline{\mc M}_{g,n+1}\to\overline{\mc M}_{g,n}$ that forgets the $(n{+}1)$-th marked point; see Appendix A in \cite{do2011} for more details.}

Physically, $2\pi^2\kappa_1$ is (a multiple of) the Weil--Petersson K\"ahler form, 
so $\int\exp(2\pi^2\kappa_1)$ recovers the classical Weil--Petersson volume. Eq.~(\ref{6.1}) is the natural $q$-deformation 
of this statement, with $\Omega_q$ replacing $2\pi^2\kappa_1$ and the $\psi_i$-term accounting for the boundary lengths $L_i$.

The quantities $s_{m}(q)$ are defined through the Gaussian-moment generating function
\be
\la{6.2}
\exp\Big(-\sum_{m\geq1}s_{m}(q)\l^{m}\Big) = \frac1{\sqrt{2\pi\l^{3}}}\int_{-\infty}^{\infty}dz\,z^{2}\exp\Big(-\sum_{m\geq1}\frac{\zeta_{q}(2m)}{m}(4z^{2})^{m}\Big)e^{-z^{2}/(2\l)}.
\ee
The first three read \cite{DoNorbury:2025dssyk}
\be
\la{6.3}
s_{1}=12\zeta_{q}(2),\quad s_{2}=24\big(5\zeta_{q}(4)-2\zeta_{q}(2)^{2}\big),\quad s_{3}=64\big(35\zeta_{q}(6)-30\zeta_{q}(2)\zeta_{q}(4)+4\zeta_{q}(2)^{3}\big),
\ee
and expanding \eqref{6.2} one order further gives the fourth coefficient
\be
\la{6.4}
s_{4}(q) = -1536\,\zeta_{q}(2)^{4}+23040\,\zeta_{q}(2)^{2}\zeta_{q}(4)-53760\,\zeta_{q}(2)\zeta_{q}(6)-23040\,\zeta_{q}(4)^{2}+60480\,\zeta_{q}(8),
\ee
and so on. In the limit $q\to1$, $s_{1}\to2\pi^{2}$ and $s_{m\geq2}\to0$, so \eqref{6.1} reduces to the classical Mirzakhani--Zograf 
statement $V^{\rm WP}_{g,n}=\int\exp(2\pi^{2}\kappa_{1})$ \cite{Mirzakhani:2006fta,Mirzakhani:2006eta}.

\subsection{Universal non-perturbative ratios}

Only $E_{2}$ carries a transformation anomaly under $\tau\to-1/\tau$; $E_{4}$ and $E_{6}$ are genuinely modular. 
Hence their exact small-$\l$ expansions, up to $O(\wt q^{2})$, are
\be
\la{6.5}
E_{2} = -\frac{4\pi^{2}}{\l^{2}}+\frac{12}\l+\frac{96\pi^{2}}{\l^{2}}\,\wt q+\cdots,
\quad
E_{4} = \frac{16\pi^{4}}{\l^{4}}\big(1+240\,\wt q\big)+\cdots,
\quad
E_{6} = -\frac{64\pi^{6}}{\l^{6}}\big(1-504\,\wt q\big)+\cdots,
\ee
and so on, where $\wt q\equiv e^{-4\pi^{2}/\l}$, introduced in (\ref{3.10}), is the non-perturbative parameter. Since every $\zeta_{q}(2k)$ 
belongs to the quasimodular ring generated by $E_{2},\dots,E_{2k}$, and its leading small-$\l$ term dominates over the lower $E_{2j}$'s, 
the leading non-perturbative correction to $\zeta_{q}(2k)$ is simply proportional to its own leading perturbative term,
\be
\la{6.6}
\zeta_{q}(2k) = \frac{\zeta(2k)}{\l^{2k}}(1+r_{k}\,\wt q)+O(\l^{-2k+1},\l^{-2k+1}\wt q,\wt q^2),
\qquad
r_{k} = -\frac{4k}{B_{2k}}.
\ee
Explicitly, $r_{k}$ is the first nonzero Fourier coefficient of the ordinary weight-$2k$ Eisenstein series:
\be
\la{6.7}
r_{1}=-24,\qquad r_{2}=240,\qquad r_{3}=-504,\qquad r_{4}=480, \qquad \dots .
\ee

\subsection{The shift operator $\mc R$}

A product $\prod_{i}\zeta_{q}(2k_{i})$ receives its $O(\wt q)$ piece from exactly one factor at a time. 
Hence the leading non-perturbative correction to any polynomial $P\big(\zeta_{q}(2),\zeta_{q}(4),\dots\big)$ is obtained by applying the fixed, $(g,n)$-independent operator
\be
\la{6.8}
\mc R \equiv \sum_{k\geq1} r_{k}\,\zeta_{q}(2k)\,\partial_{\zeta_{q}(2k)}
\ee
to $P$ and evaluating at the classical zeta values, $\zeta_{q}(2k)\to\zeta(2k)$. Applying $\mc R$ to $s_{m}$ (see \eg \eqref{6.3}, \eqref{6.4}) gives universal numbers, needed once and for all,
\be
\la{6.9}
\delta s_{1}\equiv(\mc R\,s_{1})\big|_{\rm cl} = -48\pi^{2},\qquad
\delta s_{2} = 384\pi^{4},\qquad
\delta s_{3} = -2048\pi^{6},\qquad
\delta s_{4} = 8192\pi^{8}, \qquad \dots .
\ee
The universal non-perturbative shift of the $\kappa$-potential coefficients, $\delta s_{m}\equiv(\mc R\,s_{m})|_{\rm cl}$, 
is derived in Appendix~\ref{app:deltas}, and one gets
\be
\la{6.10}
\delta s_{m} = \frac{3\,(-16\pi^{2})^{m}}{m!},
\qquad\text{equivalently}\qquad
\sum_{m\geq1}\delta s_{m}\,\l^{m} = 3\big(e^{-16\pi^{2}\l}-1\big).
\ee

\subsection{An example: $(g,n)=(2,1)$}

The $L=0$ constant term of $V^{q}_{2,1}$ is \cite{DoNorbury:2025dssyk}
\be
\la{6.11}
V^{q}_{2,1}(0) = \frac{845}{12}\zeta_{q}(2)^{4}+\frac{399}2\zeta_{q}(2)^{2}\zeta_{q}(4)+\frac{185}4\zeta_{q}(4)^{2}+\frac{406}3\zeta_{q}(2)\zeta_{q}(6)+\frac{105}2\zeta_{q}(8),
\ee
with classical limit $V^{\rm WP}_{2,1}(0)=\frac{29\pi^{8}}{192}$. Applying \eqref{6.6}, \eqref{6.7} directly to \eqref{6.11} gives the leading non-perturbative correction
\be
\la{6.12}
V^{q}_{2,1}(0) = \frac{29\pi^{8}}{192\,\l^{8}} + \cdots \ -\ \frac{53\pi^{8}}{90\,\l^{8}}\,e^{-4\pi^{2}/\l}+\cdots.
\ee
As a nontrivial check, the same number can be reproduced through the $\kappa$-classes: matching \eqref{6.11} 
against $\int_{\overline{\mc M}_{2,1}}\exp\big(s_{1}\kappa_{1}+s_{2}\kappa_{2}+s_{3}\kappa_{3}+s_{4}\kappa_{4}\big)$, truncated to degree $4=\dim\overline{\mc M}_{2,1}$, determines
the intersection numbers\footnote{
The first entry matches $V^{\rm WP}_{2,1}(0)\cdot4!/(2\pi^{2})^{4}$; the other four are new information, invisible to the classical limit since $s_{2},s_{3},s_{4}\to0$ as $q\to1$.
}
\be
\la{6.13}
\int\kappa_{1}^{4}=\frac{29}{128},\quad
\int\kappa_{1}^{2}\kappa_{2}=\frac{259}{5760},\quad
\int\kappa_{2}^{2}=\frac{53}{5760},\quad
\int\kappa_{1}\kappa_{3}=\frac{13}{1920},\quad
\int\kappa_{4}=\frac1{1152}.
\ee
Then,
\be
\la{6.14}
\delta s_{1}\,\frac{s_{1}^{3}}6\int\kappa_{1}^{4}
+\delta s_{2}\,\frac{s_{1}^{2}}2\int\kappa_{1}^{2}\kappa_{2}
+\delta s_{3}\,s_{1}\int\kappa_{1}\kappa_{3}
+\delta s_{4}\int\kappa_{4}
\ \Big|_{s_{1}=2\pi^{2}}
= -\frac{53\pi^{8}}{90},
\ee
in exact agreement with \eqref{6.12}.

\subsection{General formula}

Combining the above, we find that the leading non-perturbative correction to any $V^{q}_{g,n}(0)$ is
\be
\la{6.15}
[V^{q}_{g,n}(0)]_{O(\wt q)}^{\rm LO} = \l^{-(6g-6+2n)}\,e^{-4\pi^{2}/\l}\sum_{m=1}^{3g-3+n}\delta s_{m}\;\partial_{s_{m}}
\Big[\int_{\overline{\mc M}_{g,n}}\exp\Big(\sum_{m'\geq1}s_{m'}\kappa_{m'}\Big)\Big]_{\substack{s_{1}=2\pi^{2}\\ s_{m'\geq2}=0}}+\cdots,
\ee
and this is fixed entirely by (i) the universal numbers \eqref{6.9} (extendable to $\delta s_{5},\delta s_{6},\dots$ by \eqref{6.10}) and (ii) ordinary $\kappa$-class 
intersection numbers on $\overline{\mc M}_{g,n}$, with no further $q$-dependent input.\footnote{Note that the object being differentiated depends on all $s_{m'}$; 
the restriction $s_{1}=2\pi^{2},\,s_{m'\geq2}=0$ is imposed only after differentiating -- setting it before would trivially 
kill every $m\geq2$ term.}
Since $d=3g-3+n$, dimension forces every other $s_{m'}$-derivative to vanish at this point except the one pairing $\kappa_m$ with the remaining top power of $\kappa_1$,
\be
\la{6.16}
\partial_{s_{m}}\Big[\int_{\overline{\mc M}_{g,n}}\exp\Big(\sum_{m'\geq1}s_{m'}\kappa_{m'}\Big)\Big]_{\substack{s_{1}
=2\pi^{2}\\ s_{m'\geq2}=0}} = \frac{(2\pi^{2})^{d-m}}{(d-m)!}\int_{\overline{\mc M}_{g,n}}\kappa_{m}\kappa_{1}^{d-m} .
\ee
Substituting \eqref{6.16} into \eqref{6.15}, then substituting the closed form for $\delta s_m$ from \eqref{6.10} 
-- using $(-16\pi^2)^m(2\pi^2)^{d-m}=(2\pi^2)^d(-8)^m$ and $\frac1{m!(d-m)!}=\frac1{d!}\binom dm$ -- 
and combining $(2\pi^2)^d$ with $\l^{-(6g-6+2n)}=\l^{-2d}$ into $(2\pi^2/\l^2)^d$, gives the final, fully explicit formula,
\be
\la{6.17}
\boxed{
[V^{q}_{g,n}(0)]_{O(\wt q)}^{\rm LO} = \frac{3}{d!}\left(\frac{2\pi^{2}}{\l^{2}}\right)^{d}e^{-4\pi^{2}/\l}
\sum_{m=1}^{d}\binom{d}{m}(-8)^{m}\int_{\overline{\mc M}_{g,n}}\kappa_{m}\kappa_{1}^{d-m}+\cdots,
}
\ee
a finite sum over $m=1,\dots,d$ of ordinary $\kappa_{m}\kappa_{1}^{d-m}$-intersection numbers with purely numerical (rational) binomial weights 
-- no further $q$-dependent input beyond the overall prefactor is needed anywhere. The combination $2\pi^2/\l^2$ is recognizable as the leading small-$\l$ term 
of $s_1(q)=12\zeta_q(2)$ itself (cf.\ \eqref{6.6} at $k=1$), so \eqref{6.17} reads naturally as ``$s_1^{d}$'' at its classical leading value, reweighted by the binomial/$\kappa$-intersection sum.
An explicit algorithm to evaluate the intersection integrals is reviewed in full detail in Appendix~\ref{app:DVV}.

It is worth contrasting this with the discrete recursion of \cite{Giacchetto:2025nnb}, which determines the 
$N_{g,n}(b)$ themselves. That recursion is exact in $q$ and therefore contains the non-perturbative information 
implicitly, but extracting it requires solving for the full volume at the given $(g,n)$ and only then expanding at 
small $\l$. The content of \eqref{6.17} is that the leading non-perturbative piece can be written down directly, as 
a finite sum of intersection numbers, without that intermediate step -- which is what makes cases such as $(4,1)$ 
accessible.

\subsection{Explicit results for various $(g,n)$}
\la{sec:table}

Since $\int_{\overline{\mc M}_{g,n}}\kappa_{1}^{j}$ vanishes unless $j=d\equiv D/2=3g-3+n$ 
(top degree, $\dim\overline{\mc M}_{g,n}$), the $\delta s_{1}$-term in \eqref{6.15} collapses to a pure rescaling of the already-known classical volume,
\be
\la{6.18}
\big[\delta s_{1}\text{-term}\big] = \delta s_{1}\,\frac{d}{2\pi^{2}}\,V^{\rm WP}_{g,n}(0) = -24\,d\;V^{\rm WP}_{g,n}(0),
\ee
with all further $(g,n)$-dependence carried by the $\kappa_{\geq2}$-intersection remainder. 
Evaluating \eqref{6.6}, \eqref{6.7} on every $(g,n)$ for which $V^{q}_{g,n}(0)$ is given explicitly in \cite{DoNorbury:2025dssyk}'s own appendix table gives, in every case,

\be
\la{6.19}
V^{q}_{g,n}(0) = \Big(\frac\pi\l\Big)^{2d}[\sssa_{g,n}+\cdots+\sssb_{g,n}\,\wt q+\cdots], \qquad d=3g-3+n,
\ee
where $\sssa_{g,n}=V^{\rm WP}_{g,n}(0)/\pi^{2d}$ is the (rational) classical Weil--Petersson coefficient and $\sssb_{g,n}$ is the 
leading non-perturbative coefficient from \eqref{6.15}; ``$\cdots$'' denotes terms subleading in $1/\l$, of either kind. Explicit values of the coefficients 
$\sssa_{g,n}$ and $\sssb_{g,n}$ are collected in Table~\ref{tab:322a}.

\begin{table}[htbp]
\centering
\def\arraystretch{1.3}
\begin{tabular}{cccc}
\toprule
$(g,n)$ & $d$ & $\sssa_{g,n}$ & $\sssb_{g,n}$ \\
\midrule
$(0,4)$ & $1$ & $2$ & $-48$ \\
$(1,1)$ & $1$ & $\frac1{12}$ & $-2$ \\
$(0,5)$ & $2$ & $10$ & $-96$ \\
$(1,2)$ & $2$ & $\frac14$ & $4$ \\
$(0,6)$ & $3$ & $\frac{244}3$ & $-992$ \\
$(2,1)$ & $4$ & $\frac{29}{192}$ & $-\frac{53}{90}$ \\
$(2,2)$ & $5$ & $\frac{787}{480}$ & $-\frac{2183}{180}$ \\
\bottomrule
\end{tabular}
\caption{Classical ($\sssa_{g,n}$) and leading non-perturbative ($\sssb_{g,n}$) coefficients for $V^q_{g,n}(0)$, in the normalization of \eqref{6.19}, for every $(g,n)$ given explicitly in \cite{DoNorbury:2025dssyk}. 
The $(0,6)$ and $(2,2)$ entries are computed from our own corrected expressions for $V^q_{g,n}(0)$, not from the ones in 
\cite{DoNorbury:2025dssyk}'s appendix table.\protect\footnotemark}
\la{tab:322a}
\end{table}
\footnotetext{For these two $(g,n)$ our expressions for $V^q_{g,n}(0)$ differ from those tabulated in 
the appendix of \cite{DoNorbury:2025dssyk}; we thank N.~Do and P.~Norbury for correspondence confirming that the tabulated 
expressions in v1 of their paper contain typographical errors and that ours are the correct ones. The difference is invisible in the $q\to1$ limit -- 
both give the correct Weil--Petersson volumes -- and appears only in the non-perturbative sector, where the 
tabulated expressions would give $\sssb_{0,6}=-7904$ and $\sssb_{2,2}=-\tfrac{7609}{12}$; 
see Appendix~\ref{app:discrepancy}.}
This agrees with the direct calculation of Section~\ref{sec:N11-N12} (eq.~\eqref{4.5}) once the normalization factor $2^{3-2g-n}$ of \eqref{5.1} is taken into account.

\section{Extension to nonzero boundary lengths $L_{i}$}
\la{sec:nonzero}

\subsection{Correction in terms of mixed $\kappa$--$\psi$ integrals}

Recall from Section~\ref{sec:N11-N12} that the superscript ``LO'' denotes the leading small-$\l$ term at $O(\wt q)$, 
applied there only to the $L=0$ constant term of $V^q_{g,n}(L)$. By the same degree argument used in Section~\ref{sec:bridge}, 
each coefficient of $\prod_iL_i^{2\alpha_i}$ is itself homogeneous of a definite degree, $D-2|\alpha|$, and hence has its own leading small-$\l$ term 
-- generally at a different power of $\l$ than the $L=0$ case. Making this precise is the content of this section.

Since the boundary lengths $L_{i}$ carry no $q$-dependence, we may write
\be
\exp\Big(\sum_{m}s_{m}\kappa_{m}+\frac12\sum_{i}L_{i}^{2}\psi_{i}\Big)=\exp\Big(\sum_{m}s_{m}\kappa_{m}\Big)\ \prod_{i}\exp\Big(\frac{L_{i}^{2}}2\psi_{i}\Big).
\ee
The leading non-perturbative shift then acts only on the $\kappa$-part, while the $\psi_{i}^{\alpha_{i}}$ factors are spectators. 
Differentiating at the classical point exactly as in \eqref{6.16}, but now retaining the $\psi_{i}^{\alpha_{i}}$ 
insertions that the $L=0$ case discarded, and extracting the coefficient of one specific monomial $\prod_iL_i^{2\alpha_i}$ at a time (with $j=d-|\alpha|-m$ then fixed, rather than independently summed),
gives
\ba
\la{7.2}
[\text{coeff.\ of } & \textstyle\prod_iL_i^{2\alpha_i}\text{ in }V^{q}_{g,n}(L)]_{O(\wt q)}^{\rm LO} \lp
= \frac{\l^{-(D-2|\alpha|)}e^{-4\pi^{2}/\l}}{\prod_i2^{\alpha_i}\alpha_i!}\sum_{m=1}^{d-|\alpha|}\delta s_{m}
\,\frac{(2\pi^{2})^{d-|\alpha|-m}}{(d-|\alpha|-m)!}\int_{\overline{\mc M}_{g,n}}\kappa_{m}\kappa_{1}^{d-|\alpha|-m}\prod_i\psi_{i}^{\alpha_{i}},
\ea
with $d=D/2=3g-3+n$ as before. Setting $\alpha=0$ recovers \eqref{6.17} exactly; \eqref{7.2} is the same statement with the $\psi$-insertions restored. 

The only new ingredient is $\int\kappa_{m}\kappa_{1}^{j}\prod_{i}\psi_{i}^{\alpha_{i}}$ in place of the pure $\kappa$-numbers used in 
Appendix~\ref{app:DVV}. This requires no new algorithm: the $\kappa$-removal recursion \eqref{C.6} always 
terminates on $\langle\tau_{0}^{n}\cdots\rangle$ by convention -- allowing the $n$ original marked points to carry $\tau_{\alpha_{i}}$ 
instead of $\tau_{0}$ is a one-line generalization of the base case, with \eqref{C.4} unchanged.

Let us note two general features of \eqref{7.2}. First, at $d=1$ (i.e.\ $(0,4)$ and $(1,1)$) the $L_i^2$ coefficient is a pure topological number 
with no $\kappa$-insertion possible at all ($\alpha_i=1$ already saturates $d$), hence carries no $\zeta_q$-dependence 
and no non-perturbative correction at any order: $V^q_{0,4}(L)=\frac12\sum_iL_i^2+12\zeta_q(2)$ and $V^q_{1,1}(L)=\frac1{48}L_1^2+\frac12\zeta_q(2)$ exactly. 
Second, for any $(g,n)$ the top-degree-in-$L$ monomial (using the full dimension $d$ via $\psi$-insertions alone) always has vanishing leading NP correction, 
since $j+|\alpha|=d$ then forces $m\leq0$ for every term in \eqref{7.2}.

\subsection{An example: the $L_{1}^{2}$ coefficient of $V^{q}_{2,1}(L)$}

As an illustration, let us consider the $L_{1}^{2}$ coefficient of $V^{q}_{2,1}(L)$ \cite{DoNorbury:2025dssyk},
\be
\la{7.3}
\frac1{24}\Big(191\,\zeta_{q}(2)^{3}+243\,\zeta_{q}(2)\zeta_{q}(4)+70\,\zeta_{q}(6)\Big).
\ee
We can compute its leading small-$\l$ correction at $O(\wt q)$ by two independent routes. The first is by application of (\ref{7.2}). 
Here $d=4$, $n=1$; the $L_{1}^{2}$ term needs $\alpha_{1}=1$, so $j=3-m$ for $m=1,2,3$ ($m=4$ leaves no room for $\alpha_{1}=1$ and drops out). 
Using the methods in Appendix~\ref{app:DVV}, we have
\be
\la{7.4}
\int\kappa_{1}^{3}\psi_{1}=\frac{169}{1920},\qquad
\int\kappa_{1}\kappa_{2}\psi_{1}=\frac{101}{5760},\qquad
\int\kappa_{3}\psi_{1}=\frac1{384},
\ee
and assembling via \eqref{7.2} with the closed-form $\delta s_{1},\delta s_{2},\delta s_{3}$ of \eqref{6.10} gives
\be
\la{7.5}
-\frac{19\pi^{6}}{120}.
\ee
Since we know the explicit expression (\ref{7.3}), we can also obtain the same by applying the shift operator $\mc R$ of \eqref{6.8} straight to \eqref{7.3} 
and evaluating at the classical $\zeta(2k)$,
\be
\la{7.6}
\mc R\Big[\tfrac1{24}(191\zeta_{q}(2)^{3}+243\zeta_{q}(2)\zeta_{q}(4)+70\zeta_{q}(6))\Big]\Big|_{\rm cl} = -\frac{19\pi^{6}}{120}.
\ee

\subsection{Extended explicit results for the tabulated $(g,n)$}
\la{sec:extended-DN}

Applying \eqref{7.2} systematically (with mixed $\kappa$-$\psi$ integrals computed as in Appendix~\ref{app:DVV}), we find that
every coefficient of $\prod_iL_i^{2\alpha_i}$ in $V^q_{g,n}(L)$ takes the same form as \eqref{6.19}, with $d$ replaced by $d-|\alpha|$,
\be
\la{7.7}
\Big[\text{coeff.\ of }\textstyle\prod_iL_i^{2\alpha_i}\text{ in }V^q_{g,n}(L)\Big] = \Big(\frac\pi\l\Big)^{2(d-|\alpha|)}\Big[\sssa_{g,n,\alpha}+\cdots+\sssb_{g,n,\alpha}\,\wt q+\cdots\Big],
\ee
where $\sssa_{g,n,\alpha}$ is the classical coefficient and $\sssb_{g,n,\alpha}$ the leading non-perturbative coefficient from \eqref{7.2}; 
a dash in the $\sssa_{g,n,\alpha}$ column means the classical coefficient is the standard value, omitted since 
only $\sssb_{g,n,\alpha}$ is new; see Table~\ref{tab:47a}. The $(2,2)$ case (right panel) is now complete: by the $L_1\leftrightarrow L_2$ 
symmetry of $(2,2)$, it covers every monomial in $L_1^2,L_2^2$ up to and including top degree.

\begin{table}[htb]
\centering
\def\arraystretch{1.3}
\begin{minipage}[t]{0.52\textwidth}
\vspace{0pt}
\centering
\begin{tabular}{ccccc}
\toprule
$(g,n)$ & mon. & $d{-}|\alpha|$ & $\sssa$ & $\sssb$ \\
\midrule
$(0,5)$ & $1$ & $2$ & $10$ & $-96$ \\
$(0,5)$ & $L_i^2$ & $1$ & $3$ & $-72$ \\
\midrule
$(1,2)$ & $1$ & $2$ & $\frac14$ & $4$ \\
$(1,2)$ & $L_i^2$ & $1$ & $1$ & $-2$ \\
\midrule
$(0,6)$ & $1$ & $3$ & $\frac{244}3$ & $-992$ \\
$(0,6)$ & $L_i^2$ & $2$ & \text{--} & $-480$ \\
$(0,6)$ & $L_i^4$ & $1$ & \text{--} & $-36$ \\
$(0,6)$ & $L_i^2L_j^2$ & $1$ & \text{--} & $-144$ \\
\midrule
$(2,1)$ & $1$ & $4$ & $\frac{29}{192}$ & $-\frac{53}{90}$ \\
$(2,1)$ & $L_1^2$ & $3$ & \text{--} & $-\frac{19}{120}$ \\
$(2,1)$ & $L_1^4$ & $2$ & \text{--} & $-\frac{23}{480}$ \\
$(2,1)$ & $L_1^6$ & $1$ & \text{--} & $-\frac{29}{5760}$ \\
\bottomrule
\end{tabular}
\end{minipage}%
\begin{minipage}[t]{0.48\textwidth}
\vspace{0pt}
\centering
\begin{tabular}{ccccc}
\toprule
$(g,n)$ & mon. & $d{-}|\alpha|$ & $\sssa$ & $\sssb$ \\
\midrule
$(2,2)$ & $1$ & $5$ & $\frac{787}{480}$ & $-\frac{2183}{180}$ \\
$(2,2)$ & $L_1^2$ & $4$ & \text{--} & $-\frac92$ \\
$(2,2)$ & $L_1^4$ & $3$ & $\frac{551}{8640}$ & $-\frac{221}{360}$ \\
$(2,2)$ & $L_1^2L_2^2$ & $3$ & $\frac7{36}$ & $-2$ \\
$(2,2)$ & $L_1^6$ & $2$ & $\frac{19}{7680}$ & $-\frac{11}{288}$ \\
$(2,2)$ & $L_1^4L_2^2$ & $2$ & $\frac{181}{11520}$ & $-\frac{13}{48}$ \\
$(2,2)$ & $L_1^8$ & $1$ & $\frac{11}{276480}$ & $-\frac{11}{11520}$ \\
$(2,2)$ & $L_1^6L_2^2$ & $1$ & $\frac{29}{69120}$ & $-\frac{29}{2880}$ \\
$(2,2)$ & $L_1^4L_2^4$ & $1$ & $\frac7{7680}$ & $-\frac7{320}$ \\
\bottomrule
\end{tabular}
\end{minipage}
\caption{Classical ($\sssa\equiv \sssa_{g,n,\alpha}$) and leading non-perturbative ($\sssb\equiv \sssb_{g,n,\alpha}$) coefficients for the $L$-dependent extension \eqref{7.7}, 
for every computed monomial $\prod_iL_i^{2\alpha_i}$: $(0,5),(1,2),(0,6),(2,1)$ on the left, $(2,2)$ on the right. Top-degree-in-$L$ monomials
 (vanishing $\sssb_{g,n,\alpha}$ at every order, by the general argument above) are omitted throughout -- e.g.\ for $(2,2)$: $L_1^{10},L_1^8L_2^2,L_1^6L_2^4$, 
 with classical coefficients $\frac1{4423680},\frac1{294912},\frac{29}{2211840}$ respectively, matching Appendix A in \cite{DoNorbury:2025dssyk}.}
\la{tab:47a}
\end{table}

\section{Predictions for $(3,1)$, $(3,2)$, and $(4,1)$}
\la{sec:beyond-DN}

Every result so far has concerned a $(g,n)$ for which the  $q$-deformed volume $V^q_{g,n}(L)$ is already given explicitly in \cite{DoNorbury:2025dssyk}'s 
own appendix -- itself obtained by first running Mirzakhani's classical recursion, then the $q$-deformed recursion of \cite{DoNorbury:2025qmirz}, to get $V^q_{g,n}(L)$ in closed form.

We now consider a novel case, $(g,n)=(3,1)$, which we treat by our algorithm based on \eqref{6.17} and \eqref{7.2}. For $L=0$, 
we need only $\int\kappa_m\kappa_1^{d-m}$, computable directly via the algorithm of Appendix~\ref{app:DVV}; neither topological recursion nor any prior knowledge of $V^q_{g,n}(L)$ is required at all. 
For the $L$-dependent terms, we will need, in addition, mixed $\kappa$-$\psi$ intersection numbers $\int\kappa_m\kappa_1^{d-1-m}\psi_1$ on $\overline{\mc M}_{3,1}$, via \eqref{7.2}
-- again computable directly via the algorithm of Appendix~\ref{app:DVV}, with no new input required.

Here $d=3g-3+n=7$, so \eqref{6.17} requires the seven numbers $\int_{\overline{\mc M}_{3,1}}\kappa_m\kappa_1^{7-m}$, $m=1,\dots,7$, computed exactly as for every other $(g,n)$ above, via the algorithm of Appendix~\ref{app:DVV}:
\be
\la{8.1}
\begin{gathered}
\int\kappa_1^7 = \frac{9292841}{103680}, \qquad \int\kappa_1^5\kappa_2 = \frac{435985}{41472}, \qquad \int\kappa_1^4\kappa_3 = \frac{382073}{414720}, \qquad \int\kappa_1^3\kappa_4 = \frac{96587}{1451520}, \\
\int\kappa_1^2\kappa_5 = \frac{6049}{1451520}, \qquad \int\kappa_1\kappa_6 = \frac{97}{414720}, \qquad \int\kappa_7 = \frac1{82944}.
\end{gathered}
\ee
The first of these gives the classical coefficient directly, $\sssa_{3,1}=2^7\int\kappa_1^7/7!=\frac{9292841}{4082400}$.
This matches the value $\frac{9292841\,\pi^{14}}{4082400}$ tabulated explicitly in \cite{do2011}'s Appendix B.
Substituting \eqref{8.1} into \eqref{6.17} gives the leading non-perturbative correction,
\be
\la{8.2}
V^q_{3,1}(0) = \Big(\frac\pi\l\Big)^{14}\Big[\frac{9292841}{4082400}+\cdots-\frac{167843}{6804}\,\wt q+\cdots\Big].
\ee
As a further illustration, let us consider the $L_1^2$ coefficient of $V^q_{3,1}(L)$. Here $\alpha=1$, so $d-|\alpha|=6$, and \eqref{7.2} requires the six numbers
\be
\la{8.3}
\begin{gathered}
\int\kappa_1^6\psi_1 = \frac{8497697}{414720}, \qquad \int\kappa_1^4\kappa_2\psi_1 = \frac{7119061}{2903040}, \qquad \int\kappa_1^3\kappa_3\psi_1 = \frac{640979}{2903040}, \\
\int\kappa_1^2\kappa_4\psi_1 = \frac{47647}{2903040}, \qquad \int\kappa_1\kappa_5\psi_1 = \frac{437}{414720}, \qquad \int\kappa_6\psi_1 = \frac5{82944}.
\end{gathered}
\ee
The first of these gives the classical coefficient, $\sssa_{3,1,\alpha=1}=2^6\int\kappa_1^6\psi_1/(6!\cdot2)=\frac{8497697}{9331200}$, 
and assembling \eqref{8.3} via \eqref{7.2} gives the leading non-perturbative correction to the same coefficient,
\be
\la{8.4}
\Big[\text{coeff.\ of }L_1^2\text{ in }V^q_{3,1}(L)\Big] = \Big(\frac\pi\l\Big)^{12}\Big[\frac{8497697}{9331200}+\cdots-\frac{13291849}{1360800}\,\wt q+\cdots\Big].
\ee
As with the $L=0$ case above, the classical coefficient in \eqref{8.4} also matches Appendix B in \cite{do2011}.
The accompanying non-perturbative coefficient, and the underlying intersection numbers in \eqref{8.3}, are new.

The same method extends directly to further cases beyond those tabulated in \cite{DoNorbury:2025dssyk}. Presented here at $L=0$ only, one finds for $(g,n)=(3,2)$ and $(4,1)$
\be
\la{8.5}
V^q_{3,2}(0) = \Big(\frac\pi\l\Big)^{16}\Big[\frac{2800144027}{65318400}+\cdots-\frac{198679583}{340200}\,\wt q+\cdots\Big],
\ee
\be
\la{8.6}
V^q_{4,1}(0) = \Big(\frac\pi\l\Big)^{20}\Big[\frac{92480712720869}{987614208000}+\cdots-\frac{10658799618671}{6858432000}\,\wt q+\cdots\Big].
\ee

\section{Beyond leading-order: the $O(\wt q^2)$ correction}
\la{sec:NP2}

Everything so far has been linear in $\wt q$. It is natural to ask whether the same construction extends to the next order, $O(\wt q^2)$ -- restricting to $L_i=0$ for simplicity -- 
and again asking only for the leading small-$\l$ term multiplying $\wt q^2$.

\subsection{The extended formula}

Recall that $\zeta_q(2k)$'s leading-$\l$ piece is controlled by a single Eisenstein series, $\zeta_q(2k)\sim\zeta(2k)E_{2k}(\wt q)/\l^{2k}$, 
with mixing from other $E_{2j}$'s entering only at subleading $\l$-order. This statement holds for the entire $q$-expansion of $E_{2k}$, 
not just its linear term -- so the leading-$\l$ piece of $\zeta_q(2k)$, to any order in $\wt q$, is obtained by simply reading off the corresponding 
Taylor coefficient of $E_{2k}(\wt q)$ itself. Concretely, writing $\wh s_m(\wt q)\equiv s_m\big|_{\zeta_q(2k)\to\zeta(2k)E_{2k}(\wt q)}$ 
for the leading-$\l$ piece of $s_m(q)$ as a function of $\wt q$ alone, we have
\be
\la{9.1}
\wh s_m(\wt q) = s_m^{\rm cl}+\delta s_m\,\wt q+\delta s_m^{(2)}\,\wt q^2+O(\wt q^3),
\ee
with $\delta s_m$ the already-known linear-order shift and $\delta s_m^{(2)}\equiv\tfrac12\wh s_m''(0)$ the new quantity of interest.

Using the second Fourier coefficients of $E_2,E_4,E_6$ -- $-72,\,2160,\,-16632$ respectively -- and Taylor-expanding \eqref{9.1} 
directly from the known formulas for $s_1,\ldots,s_4$ (eqs.~\eqref{6.3}, \eqref{6.4}) gives, for $m=1,2,3,4$,
\be
\la{9.2}
\delta s_1^{(2)}=-144\pi^2,\qquad \delta s_2^{(2)}=2304\pi^4,\qquad \delta s_3^{(2)}=-24576\pi^6,\qquad \delta s_4^{(2)}=196608\pi^8.
\ee
All four match, exactly, the closed form
\be
\la{9.3}
\delta s_m^{(2)} = \frac92\,\frac{(-32\pi^2)^m}{m!}, \qquad\text{equivalently}\qquad \sum_{m\geq1}\delta s_m^{(2)}\l^m = \frac92\big(e^{-32\pi^2\l}-1\big).
\ee
This is the same shape as $\delta s_m=3(-16\pi^2)^m/m!$ (eq.~\eqref{6.10}), with the exponent doubled and the prefactor scaled by $3/2$; 
we give a full derivation, extending the generating-function argument used for $\delta s_m$ to second order, in Appendix~\ref{app:deltas}.

Expanding $\exp(\Omega_q)=\exp(\sum_m s_m\kappa_m)$ to $O(\wt q^2)$ using \eqref{9.1} gives, beyond the classical piece,
\be
\la{9.4}
\exp(\Omega^{\rm cl})\times\bigg[\underbrace{\sum_m\delta s_m^{(2)}\kappa_m}_{\text{new}}+\underbrace{\frac12\Big(\sum_m\delta s_m\kappa_m\Big)^2}_{\text{cross term}}\bigg]\wt q^2+O(\wt q^3).
\ee
This requires mixed, two-insertion intersection numbers $\int\kappa_m\kappa_{m'}\kappa_1^{d-m-m'}$ that never appeared at $O(\wt q)$, 
but which are computable with the same DVV/$\kappa$-removal algorithm of Appendix~\ref{app:DVV}, applied to a different insertion pattern.

Combining \eqref{9.4} with the same dimension-counting argument as \eqref{6.17} gives, for the leading small-$\l$ correction at $O(\wt q^2)$,
\ba
\la{9.5}
\sssc_{g,n} &= \frac{9\, 2^{d}}{2\,d!}\bigg[\sum_{m=1}^{d}\binom{d}{m}(-16)^m\!\int\!\kappa_m\kappa_1^{d-m} \;+\; \lp
\sum_{\substack{m,m'\geq1\\ m+m'\leq d}}\binom{d}{m,m',d-m-m'}(-8)^{m+m'}\!\int\!\kappa_m\kappa_{m'}\kappa_1^{d-m-m'}\bigg],
\ea
with $V^q_{g,n}(0)=(\pi/\l)^{2d}[\sssa_{g,n}+\sssb_{g,n}\wt q+\sssc_{g,n}\wt q^2+\cdots]$. 
The first sum is structurally identical to \eqref{6.17} itself, with $(-8)^m\to(-16)^m$ and prefactor $3\to\tfrac92$; the second is new.

One can check that the degree-counting argument of Section~\ref{sec:bridge} still isolates the leading $O(\wt q^2)$ term correctly, 
exactly as Section~\ref{sec:N11-N12} checked this directly at $O(\wt q)$. The argument itself is unchanged in principle: a term of joint 
degree $d'$ has leading $\l$-power at most $\l^{-d'}$ at \emph{every} order in $\wt q$, since this only relies on $\zeta_q(2k)$'s own leading-$\l$ structure, 
which is $\wt q$-order-independent. We can examine this mechanism at work in explicit examples.

For $(1,1)$: $\wt N_{1,1}(b)=[\tfrac{b^2-4}{48}+\tfrac{\zeta_q(2)}2]P_b$, and only $\zeta_q(2)/2$ contributes to $O(\wt q^2)$; 
its leading-$\l$, $O(\wt q^2)$ coefficient is $-6\pi^2$, matching $2^{3-2g-n}\sssc_{1,1}\pi^2=1\times(-6)\pi^2$ exactly.

For $(1,2)$ -- the meaningful test, since here a genuine lower-degree piece is present -- $\wt N_{1,2}(0,0)$'s bracket is 
$\tfrac{5}{96}-\tfrac12\zeta_q(2)+\tfrac72\zeta_q(2)^2+\tfrac52\zeta_q(4)$. The $-\tfrac12\zeta_q(2)$ term is degree $2$, strictly below the top degree $4$; 
its own leading-$\l$ behavior is $\l^{-2}$, not $\l^{-4}$, so at the $\l^{-4}$ order specifically it contributes exactly zero, at every order in $\wt q$ 
including $O(\wt q^2)$. The full bracket's leading-$\l$, $O(\wt q^2)$ coefficient works out to $102\pi^4$, matching $2^{3-2g-n}\sssc_{1,2}\pi^4=\tfrac12\times204\times\pi^4$ exactly.

\subsection{Results for the $O(\wt q^{2})$ correction}

Applying \eqref{9.5} to every $(g,n)$ in Table~\ref{tab:322a}, together with $(g,n)=(3,1)$ from Section~\ref{sec:beyond-DN}, gives the leading $O(\wt q^2)$ coefficients collected in Table~\ref{tab:NP2}. 
For the first five entries of Table~\ref{tab:322a},
the mixed intersection numbers entering the second sum of \eqref{9.5} turn out to already be a subset of the full partition list in Table~\ref{tab:kappa}, 
so these five entries are checked both ways -- directly from the exact small-$\l$ expansion of $V^q_{g,n}(0)$ itself, and from the 
assembly formula \eqref{9.5} -- as in Table~\ref{tab:322a}.
The mixed intersection numbers for $(2,1)$ and $(2,2)$ are likewise already available, so both entries follow from \eqref{9.5} directly. 
For $(2,2)$ no independent check against a closed form is possible, since the closed form there is the one we correct in 
Table~\ref{tab:322a}; see Appendix~\ref{app:discrepancy}.
The $(3,1)$ entry required six new mixed intersection numbers not computed in Section~\ref{sec:beyond-DN} -- 
$\int\kappa_2^2\kappa_1^3,\int\kappa_2\kappa_3\kappa_1^2,\int\kappa_2\kappa_4\kappa_1,\int\kappa_2\kappa_5,\int\kappa_3^2\kappa_1,\int\kappa_3\kappa_4$ 
on $\overline{\mc M}_{3,1}$ -- computed with the same algorithm of Appendix~\ref{app:DVV}; 
the resulting sum has $21$ terms in the second piece of \eqref{9.5}. The $(3,2)$ and $(4,1)$ entries are obtained the same way, 
requiring nine and sixteen new mixed intersection numbers respectively, with $28$ and $45$ terms respectively in the corresponding sum.

\begin{table}[htbp]
\centering
\def\arraystretch{1.3}
\begin{tabular}{ccc}
\toprule
$(g,n)$ & $d$ & $\sssc_{g,n}$ \\
\midrule
$(0,4)$ & $1$ & $-144$ \\
$(1,1)$ & $1$ & $-6$ \\
$(0,5)$ & $2$ & $6624$ \\
$(1,2)$ & $2$ & $204$ \\
$(0,6)$ & $3$ & $-26016$ \\
$(2,1)$ & $4$ & $\frac{6299}{30}$ \\
$(2,2)$ & $5$ & $-\frac{2503}{60}$ \\
\midrule
$(3,1)$ & $7$ & $\frac{6399793}{56700}$ \\
$(3,2)$ & $8$ & $\frac{368030353}{113400}$ \\
$(4,1)$ & $10$ & $\frac{74590156332259}{6858432000}$ \\
\bottomrule
\end{tabular}
\caption{Leading $O(\wt q^2)$ coefficient $\sssc_{g,n}$, in the normalization $V^q_{g,n}(0)=(\pi/\l)^{2d}[\sssa_{g,n}+\sssb_{g,n}\wt q+\sssc_{g,n}\wt q^2+\cdots]$ of \eqref{9.5}. 
The first seven rows extend Table~\ref{tab:322a} to the same $(g,n)$; the last three rows are the genuinely new cases of Section~\ref{sec:beyond-DN}, beyond anything tabulated in \cite{DoNorbury:2025dssyk}.}
\la{tab:NP2}
\end{table}

\section{Comments on relation to other non-perturbative results}
\la{sec:comments}

The parameter $\wt q=e^{-4\pi^2/\l}$ studied throughout this paper is not the only non-perturbative structure appearing in the 
recent literature, and it is worth being precise about how it sits among them. Three comparisons are relevant here. The first is a 
set of recent studies of non-perturbative corrections in DSSYK with respect to the $\l\to0$ limit, which reach the same scale by 
closely related methods. The second is the semiclassical saddle structure of DSSYK, which supplies $\wt q$ with a bulk 
interpretation. The third is the standard large-genus resurgence literature for matrix models and their gravity duals, 
where a different mechanism governs a different limit -- in the genus expansion parameter rather than in $\l$.

\paragraph{Comparison with recent DSSYK non-perturbative results}

Three papers study non-per\-turbative
corrections in the small-$\l$ limit directly within DSSYK and are worth commenting on together.
The closest in spirit is \cite{Okuyama:2025fhi}, which studies the exact disk partition function (\ref{1.4})
-- the $n=1$, genus-zero seed of the construction used throughout this paper. 
By writing the measure $\mu(\theta)$ in (\ref{1.4}) via a Jacobi theta function and applying its modular $S$-transform, the small-$\l$ expansion is found to be controlled by 
our same parameter $\wt q$. This is natural since the mechanism used there 
($S$-duality of a weight-$\frac12$ theta function) is the same one underlying our own $E_2,E_4,E_6$ computation.
In the low-energy limit, the non-perturbative corrections resum, to all orders in $\wt q$, into the cube of the Dedekind eta function, with the entire $\l$-dependence captured by
\be
\la{10.1}
Z\sim \eta\big(\wt q\,e^{16\pi^2/\beta_{JT}}\big)^{3},
\ee
where $\beta_{JT}$ is the rescaled Schwarzian (JT-gravity) inverse temperature, with corrections of order $O(\wt q)=O(e^{-4\pi^2/\l})$, matching our own leading non-perturbative correction.
Ref.~\cite{Okuyama:2025fhi} also proposes a bulk reading of this structure. In the low-energy limit the corrections organize as a sum 
over $\wt q^{\,\gamma_j/8}$ with $\gamma_j=(2j+1)^2$, each term weighted by a JT partition function at a shifted dilaton boundary 
condition, suggesting that semiclassical DSSYK is dual to a superposition of such boundary conditions in JT gravity with negative 
cosmological constant. It is worth stressing that ref.~\cite{Okuyama:2025fhi} explicitly leaves open how this picture relates to the 
sine-dilaton description, which is consistent with the cautious stance we take here: the $\wt q$ corrections have a well-defined 
matrix-model meaning at every $(g,n)$, while their gravitational interpretation is settled only at the disk, and there only tentatively.

In our recent paper \cite{Beccaria:2026ndg}, we reached the same $\wt q$ scale by a more direct route: there the disk partition 
function itself is organized entirely in terms of the ring $\{E_2,E_4,E_6,(q;q)_\infty\}$, closed under the Ramanujan operator $\mc D=q\,d/dq$. 
The modular input is identical to the one used here, so the agreement of the non-perturbative scale is automatic rather than a check. 
What the present work adds is that the same transformation, applied to the $\zeta_q(2k)$ appearing as coefficients of the stable $(g,n)$ discrete 
volumes, can be evaluated in closed form and reduced to intersection numbers on $\overline{\mc M}_{g,n}$.

The pattern changes in a more instructive way once a genuinely different observable is considered. In \cite{Alfinito:2026cky} we studied Krylov spread complexity
-- dual to wormhole length in sine-dilaton gravity -- via a 5-loop semiclassical expansion at infinite temperature. 
A resummation formula for the large-time linear-growth slope $A_{1}(\l)$ was proven analytically. The accompanying non-perturbative correction takes the form
\be
\la{10.2}
A_1(\l)-A_1^{\rm pert}(\l) \sim e^{-\frac{\pi^2}{2\l}}.
\ee
The exponent in \eqref{10.2} is $\frac18$ of $4\pi^2/\l$, \ie $\wt q^{1/8}$. It is worth being careful about why this is not in tension 
with the $O(\wt q)$ corrections found above. In \eqref{10.1} the power $\wt q^{1/8}$ appears as an overall prefactor -- it is the 
$q^{1/8}$ of $\eta(\tau)^{3}=q^{1/8}\prod_{n}(1-q^{n})^{3}$, equivalently the $j=0$ term of the sum over $\gamma_j$, which 
multiplies the leading perturbative JT contribution and cancels in any ratio; the first genuine relative correction there is the $j=1$ 
term, suppressed by $\wt q^{(9-1)/8}=\wt q$. In \eqref{10.2}, by contrast, $\wt q^{1/8}$ is itself the leading relative correction to 
the perturbative series. 
The two roles are therefore structurally different, and the numerical coincidence of the exponent is not by itself evidence of a common 
mechanism. What it does indicate is sharper: $A_1(\l)$ receives a genuine relative correction at $\wt q^{1/8}$, a scale at which the 
partition function receives none, while the $O(\wt q)$ corrections present in $A_1$ sit below it. A weaker saddle is thus visible to 
complexity- and length-type observables and invisible to $Z$. A gravitational identification of this saddle is not known.

\paragraph{The winding saddles and the fixed-topology expansion}

The scale $\wt q$ has a semiclassical counterpart in the DSSYK saddle structure. In the low-temperature limit of the bilocal 
Liouville description, additional subleading saddles labelled by $k\geq1$ were identified in \cite{Berkooz:2024ifu}, and discussed 
independently in the sine-dilaton context in \cite{Blommaert:2024whf}; their reparametrizations wind $2k+1$ times around the 
thermal circle, and their fluctuations are governed by a Schwarzian theory with a conical defect of deficit angle $2\pi(2k+1)$. 
Writing $\bJ$ for the DSSYK energy scale of \cite{Berkooz:2024ifu}, the saddles associated with the lower edge of the spectrum 
have on-shell actions whose differences, in the low-temperature limit, are
\be
\la{10.3}
 I_k-I_0 \ \xrightarrow[\beta \bJ\to\infty]{}\ \frac{2\pi^2}{\l}\,k(k+1),
 \qquad\text{so that}\qquad
 e^{-(I_k-I_0)}\sim\wt q^{\,k(k+1)/2}.
\ee
In particular $k=1$ reproduces $\wt q$ itself: the exponential scale generated algebraically by the modular $S$-transformation has 
a semiclassical realization as the relative weight of the first additional winding saddle. This identification was already made in 
\cite{Beccaria:2026ndg}, at the level of the disk partition function, and is supported there by the matching 
one-loop Schwarzian contributions of \cite{Berkooz:2024ifu}.

What \eqref{10.3} adds is that the agreement extends beyond a single exponent. The winding saddles populate a \emph{sparse} tower 
indexed by the triangular numbers $1,3,6,10,\dots$, and the same tower underlies the disk-level analysis discussed above: the 
exponents $\gamma_j/8=(2j+1)^2/8$ of \cite{Okuyama:2025fhi}, measured relative to the $j=0$ term, are exactly $j(j+1)/2$. At disk 
level the modular and semiclassical routes therefore agree on the entire tower, not just on the leading exponent.

Against this, our fixed-$(g,n)$ expansion runs over the \emph{full} integer tower $\wt q,\wt q^{2},\dots$, with coefficients given 
by the $\kappa$-class intersection numbers of \eqref{6.17}, supplemented by $\psi$-class insertions for finite boundary lengths. 
The two structures therefore cannot be identified term by term: $k=1$ gives $\wt q$, but $k=2$ gives $e^{-12\pi^2/\l}=\wt q^{3}$, 
not $\wt q^{2}$. Nor is the comparison straightforward, since the saddles of \cite{Berkooz:2024ifu,Blommaert:2024whf} are 
disk-level objects evaluated at $\beta\bJ\to\infty$, whereas the $V^q_{g,n}$ are $\beta$-independent coefficients in the expansion 
\eqref{2.1}. Understanding how the sparse disk-level structure emerges from the dense fixed-topology one, after transforming to 
$Z_{g,n}$ via \eqref{2.1} and resumming over genus, remains an interesting open question.

\paragraph{Eigenvalue-tunneling resurgence: a related but different problem}

The standard resurgence literature for matrix models and their gravitational duals studies eigenvalue tunneling (ZZ-brane) instantons controlling the 
large-genus growth of a matrix model's own perturbative coefficients -- see Pasquetti--Schiappa \cite{Pasquetti:2010bps}, Gu--Mari\~no's ``peacock pattern'' program \cite{Gu:2021ize}, 
and, most relevantly, Eynard \textit{et al.}'s treatment of JT gravity \cite{Eynard:2023qdr}.
Concretely, Ref.~\cite{Eynard:2023qdr} computes, for exactly the classical ($q\to1$) spectral curve underlying this entire paper, the period $\mc A_1=\int_0^{1/4}y(x)\,dx=\frac1{4\pi^2}$, controlling how
\be
\la{10.4}
V_{g,n}\sim \mc A_1^{-(2g+n-\frac52)}\,\Gamma\Big(2g+n-\frac{5}{2}\Big)
\ee
grows as $g\to\infty$ at fixed $n$ -- the standard large-order/resurgence dictionary, in which the growth-rate base and the 
instanton action controlling the non-perturbative completion are the same number, $\mc A_1$, by construction. Their genus-counting parameter is external bookkeeping, unrelated to $\l$.\footnote{
A technically independent derivation of the same instanton is given in \cite{Hatsuda:2025cwx}, working not from matrix-model saddle points but directly from the 
Witten--Kontsevich/KdV integrable structure: the generating function of $\psi$-class intersection numbers is a $\tau$-function of the KdV hierarchy, 
and the authors construct an explicit transseries solution to this hierarchy, valid for general topological-gravity couplings $\{t_k\}$, 
with the classical Weil--Petersson case recovered at the specific point $t_k=\gamma_k$. At that point the instanton action and loop corrections 
-- computed independently, by an unrelated method, to 17 loop orders -- agree exactly with the results in \cite{Eynard:2023qdr}. 
See also \cite{Johnson:2026jbq} for a recent computation of such transseries data for $V_{g,1}(b)$ directly, 
via a combination of the Gel'fand--Dikii resolvent equation and the matrix model's string equation.}

Our own $\wt q$ lives at a single, fixed $(g,n)$ -- no sum over genus $g$ at all -- and arises instead from the $\l$-dependence of the curve itself via the quasimodular $S$-duality of $E_2,E_4,E_6$. 
These are non-perturbative completions of two genuinely different divergent series, in two different expansion variables.

\section{Summary and discussion}
\la{sec:summary}

The starting point of this paper is the observation, established in \cite{Okuyama:2023kdo}, that the discrete volumes $N_{g,n}$ governing 
the genus expansion of the ETH matrix model, \cf (\ref{1.7}) and (\ref{2.1}), 
are quasi-polynomials in the discrete boundary data $b_i$, built from the $q$-zeta values $\zeta_q(2k)$.
Physically, the discreteness of the $b_i$ is not a technical artifact: on the gravity side, the sine-dilaton dual 
of \cite{Blommaert:2024ymv} maps the periodicity of the dilaton to a discretization of the length of the 
Einstein-Rosen bridge \cite{Blommaert:2024whf}.

Each individual, fixed-genus term $N_{g,n}(b)$ has a complicated dependence on the DSSYK coupling $q=e^{-\l}$. In the sine-dilaton theory, 
the limit $\l\to0$ reproduces JT gravity, and corresponds to the leading term of a systematic semiclassical expansion.
This limit is directly accessible through the modular properties of the $\zeta_q(2k)$: these are polynomial 
combinations of the quasimodular generators $E_2,E_4,E_6$ \cite{AndrewsRose:2013,BachmannKuhn:2016}, and applying the 
$S$-duality transformation $\tau\to-1/\tau$ to each $E_{2k}$ separates their small-$\l$ behaviour into a classical, 
perturbative tower together with a further, genuinely non-perturbative correction, controlled by the quantity 
$\wt q\equiv e^{-4\pi^2/\l}$. Because this transformation acts the same way on every $\zeta_q(2k)$, its effect on any 
polynomial built from them -- in particular the discrete volumes themselves -- can be packaged into a single, universal 
derivation, independent of $(g,n)$, which we call the shift operator.

A clean handle on the leading non-perturbative correction to $\zeta_q(2k)$ is not, by itself, enough to say anything about the discrete volumes 
of the matrix model: these are quasi-polynomials in the boundary data $b_i$ with several terms of different degree, and only the term of top degree coincides 
-- up to a known, elementary normalization factor -- with the $q$-deformed Weil--Petersson volumes introduced in \cite{DoNorbury:2025qmirz,DoNorbury:2025dssyk}. 
What makes the leading non-perturbative correction tractable is a simple degree-counting argument: the terms of lower degree, which are not 
captured by the $q$-deformed volumes, are necessarily subleading in the small-$\l$ expansion, and therefore cannot contribute at leading order. 
The upshot is that the leading non-perturbative correction ($O(\wt q)$ terms at leading order in the small-$\l$ expansion) to the full discrete volume is captured entirely by the $q$-deformed Weil--Petersson volumes, 
with no residual dependence on the lower-degree terms that are otherwise left unspecified. 

It remains to make the leading correction computable in practice, for arbitrary genus and number of boundaries, 
without having to repeat the topological recursion computation that originally produced the discrete volume in the first place. 
This is achieved by combining the shift operator -- applied directly to the defining relation between $\zeta_q(2k)$ and the $q$-deformed volumes, and evaluated in closed form 
using the same modular identity underlying the small-$\l$ expansion above -- with an elementary dimension-counting argument on the relevant 
moduli space of curves: only terms linear in the shift survive, each pairing a single $\kappa_{m}$
insertion with the maximal power of $\kappa_{1}$
allowed by the dimension.

The result is that the entire leading non-perturbative correction for any $(g,n)$ -- both for the volume at vanishing boundary lengths and for every 
coefficient in its expansion in powers of the boundary lengths -- reduces to a finite sum of ordinary intersection numbers of $\kappa$-classes on 
the moduli space of curves, computable by a standard, purely combinatorial recursion with no further $q$-dependence anywhere. 
Explicit results following from this reduction, covering every case for which the $q$-deformed volume has been given explicitly in the literature, are collected in tables. 
To demonstrate that the method reaches genuinely further than this, we also compute the leading non-perturbative corrections for three new cases, $(g,n)=(3,1),(3,2),(4,1)$,
beyond the cases for which $V^q_{g,n}(L)$ is currently available in closed form.

We further extend this construction one order beyond, to the leading correction at $O(\wt q^2)$. This requires two new 
ingredients: a second-order analogue of the shift operator, again available in closed form, and a cross term, arising 
from expanding the exponential defining the volume to second order, which involves mixed, two-insertion 
$\kappa$-class intersection numbers not needed at leading order.

The same framework can in principle be extended to $O(\wt q^3)$ and beyond, at the price of progressively more complicated
multi-$\kappa$ intersection numbers. Whether the resulting series resums into closed form, as it does already at the level of 
the disk partition function itself, eq.~\eqref{10.1}, or as conjectured for a different observable (Krylov spread complexity) in \cite{Alfinito:2026cky}, is a genuinely open question for the stable $(g,n)$ volumes studied here.

We also note that going beyond leading order in $\l$ would require the lower-degree components of $\wt N_{g,n}$, which are not captured by 
$V^q_{g,n}$. Recent work expressing one-matrix-model correlators as integrals over $\overline{\mc M}_{g,n}$ at finite 
't Hooft coupling \cite{Giacchetto:2026gpq} suggests these admit an intersection-theoretic description as well, though one 
involving classes supported on the boundary of moduli space in addition to $\kappa$- and $\psi$-classes.

Finally, \eqref{6.17} suggests a question we have not pursued. The ratio 
$\sssb_{g,n}/\sssa_{g,n}$, \ie the size of the non-perturbative correction relative to the classical volume, is given by, \cf \eqref{6.16}, \eqref{6.17},
\be
\la{11.1}
\frac{\sssb_{g,n}}{\sssa_{g,n}}=3\sum_{m\geq1}\binom{d}{m}(-8)^m\,
\frac{\int_{\overline{\mc M}_{g,n}}\kappa_m\kappa_1^{d-m}}{\int_{\overline{\mc M}_{g,n}}\kappa_1^{d}},
\ee
that is, the expectation value of the shift class $\sum_m\delta s_m\kappa_m$ of \eqref{6.10} with respect to the 
measure $\exp(2\pi^2\kappa_1)$ on $\overline{\mc M}_{g,n}$ that computes the classical Weil--Petersson volume.
This is precisely the type of object for which large-genus asymptotics of tautological intersection numbers are available \cite{Mirzakhani:2011gta,Aggarwal2021}.

The sum in \eqref{11.1} is subject to severe cancellations. The $m=1$ term alone equals $-24d$, whereas the full sum is 
smaller by more than an order of magnitude: at $(g,n)=(3,1)$, where $d=7$, the individual terms range up to $\simeq550$ 
in magnitude and cancel down to $\sssb_{3,1}/\sssa_{3,1}\simeq-10.8$, a factor of fifty. Any asymptotic analysis must therefore 
control the ratios uniformly in $m$, with errors small compared to the surviving sum rather than to the individual terms.
Determining the large-genus behaviour of \eqref{11.1} would quantify the separation, emphasized in Section~\ref{sec:comments}, between
the $\l$-non-perturbative scale and the factorial large-genus growth.

\acknowledgments

We thank N.~Do and P.~Norbury, on whose results this paper builds, for helpful correspondence.
MB is supported by the INFN grant GAST. EA is supported by the MUR project GINEVRA, grant no.
2022BZYBWM.

\appendix

\section{The degree hierarchy at $(g,n)=(2,1)$}
\la{app:N21}

The examples of Section~\ref{sec:N11-N12} exhibit the degree-counting mechanism in its simplest form: $\wt N_{1,1}$ has 
a single lower-degree term, which is $\wt q$-independent, and $\wt N_{1,2}$ a single lower component. We record here the 
case $(g,n)=(2,1)$, where the top degree is $D=8$ and four distinct lower components are present, so that the full 
hierarchy claimed in Section~\ref{sec:bridge} can be seen at once.

From \cite{Okuyama:2023kdo},
\ba
\la{A.1}
\wt N_{2,1}(b) &= \Biggl[\frac{(5b^2-32)(b^2-2^2)(b^2-4^2)(b^2-6^2)}{8847360}
+\frac{(29b^2-108)(b^2-2^2)(b^2-4^2)}{92160}\zeta_q(2) \lp
+\frac{(29b^2-48)(b^2-2^2)}{1536}\zeta_q(4)
+\frac{(359b^2-624)(b^2-2^2)}{7680}\zeta_q(2)^2 \lp
+\frac{191b^2+4}{96}\zeta_q(2)^3+\frac{81b^2+28}{32}\zeta_q(2)\zeta_q(4)
+\frac{7(5b^2+4)}{48}\zeta_q(6) \lp
+\frac{845}{48}\zeta_q(2)^4+\frac{185}{16}\zeta_q(4)^2+\frac{399}8\zeta_q(2)^2\zeta_q(4)
+\frac{203}6\zeta_q(2)\zeta_q(6)+\frac{105}8\zeta_q(8)\Biggr]P_b.
\ea
Assigning $\deg b=1$ and $\deg\zeta_q(2k)=2k$ as in Section~\ref{sec:bridge}, and expanding the $b$-polynomials, 
\eqref{A.1} decomposes into components of degree $8,6,4,2$ and $0$. Substituting \eqref{3.17} into each component separately 
and retaining the term linear in $\wt q$ -- exact in $\l$ -- gives leading small-$\l$ behaviour, \cf \eqref{4.4},
\be
\la{A.2}
\big[\text{deg }8\big]:\ -\frac{53\pi^8}{360\,\l^8},\qquad
\big[\text{deg }6\big]:\ \frac{\pi^6}{40\,\l^6},\qquad
\big[\text{deg }4\big]:\ -\frac{\pi^4}{10\,\l^4},\qquad
\big[\text{deg }2\big]:\ \frac{3\pi^2}{10\,\l^2},
\ee
while the degree-$0$ component is $\wt q$-independent and drops out identically. Each component therefore contributes at 
exactly $\l^{-d'}$, with $d'$ its own degree: the bound of Section~\ref{sec:bridge} is saturated at every level rather 
than merely satisfied, and the four lower components are suppressed by $\l^2,\l^4,\l^6,\l^8$ relative to the top one. Only 
the degree-$8$ component survives at leading order, as claimed.

Retaining also the $b$-dependence of the top component gives
\be
\la{A.3}
\big[\wt N_{2,1}(b)\big]_{O(\wt q)}^{\rm LO} = -\frac{53\pi^8}{360\,\l^8}-\frac{19\pi^6}{480\,\l^6}b^2
-\frac{23\pi^4}{1920\,\l^4}b^4-\frac{29\pi^2}{23040\,\l^2}b^6,
\ee
with no $b^8$ term, in accordance with the general statement of Section~\ref{sec:nonzero} that the top-degree-in-$L$ 
monomial has vanishing leading non-perturbative correction. Using $\wt N_{g,n}=2^{3-2g-n}V^q_{g,n}$ of \eqref{5.1}, 
which for $(2,1)$ is a factor $\tfrac14$, the four coefficients in \eqref{A.3} correspond to 
$\sssb_{2,1}=-\tfrac{53}{90}$ and $\sssb_{2,1,\alpha}=-\tfrac{19}{120},-\tfrac{23}{480},-\tfrac{29}{5760}$ for the 
$L_1^2,L_1^4,L_1^6$ monomials -- exactly the values obtained from \eqref{6.17} and \eqref{7.2} in 
Tables~\ref{tab:322a} and \ref{tab:47a}. This is an independent check of the whole construction at $(2,1)$: 
\eqref{A.3} uses only the discrete volumes in  \cite{Okuyama:2023kdo} and the expansions \eqref{3.17}, with no intersection numbers and no 
reference to $V^q_{g,n}$.

\section{Derivation of the closed form of $\delta s_m$ and $\delta s_m^{(2)}$}
\la{app:deltas}

This appendix gives the derivation of the closed-form expressions for $\delta s_m$ and $\delta s_m^{(2)}$
used throughout the main text (eqs.~\eqref{6.10} and \eqref{9.3}).

These are obtained from the same generating-function identity \eqref{6.2}, but by two different routes:
$\delta s_m$ by applying the shift operator $\mc R$ directly and evaluating the resulting Gaussian integral in
closed form, while $\delta s_m^{(2)}$ instead requires substituting the full Fourier expansion of each
$\zeta_q(2k)$ and expanding to second order in $\wt q$, before an analogous Gaussian integral is evaluated.

\subsection{Proof of $\delta s_{m}$}

Since $\mc R$ is a first-order derivation acting only on the $\zeta_{q}(2k)$'s (it commutes with $\partial_{\l}$ and with the $z$-integration), 
we apply it to both sides of the defining relation \eqref{6.2}. By construction, 
$\mc R\,\zeta_{q}(2k)=r_{k}\zeta_{q}(2k)$, so we get 
\be
\la{B.1}
\mc R\,e^{-\sum_{m}s_{m}\l^{m}} = -\Big(\sum_{m}\delta_{\mc R}s_{m}\,\l^{m}\Big)e^{-\sum_{m}s_{m}\l^{m}},
\qquad
\mc R\,e^{-\sum_{m}\frac{\zeta_{q}(2m)}{m}(4z^{2})^{m}} = -G_{q}(z)\,e^{-\sum_{m}\frac{\zeta_{q}(2m)}{m}(4z^{2})^{m}},
\ee
where $\delta_{\mc R}s_{m}\equiv\mc R\,s_{m}$ (not yet evaluated at the classical point) and 
$G_{q}(z)\equiv\sum_{m}\frac{r_{m}\zeta_{q}(2m)}{m}(4z^{2})^{m}$. Equating the two sides of $\mc R$ applied to \eqref{6.2} and 
then specializing to the classical point $\zeta_{q}(2k)\to\zeta(2k)$ (so $\delta_{\mc R}s_{m}\to\delta s_{m}$, $s_{1}\to2\pi^{2}$, $s_{m\geq2}\to0$), 
and using the classical Lambert identity 
\be
\exp\big(-\sum_{m}\frac{\zeta(2m)}{m}(4z^{2})^{m}\big)=\frac{\sin(2\pi z)}{2\pi z},
\ee
gives
\be
\la{B.3}
-\Big(\sum_{m}\delta s_{m}\l^{m}\Big)e^{-2\pi^{2}\l} = \frac1{\sqrt{2\pi\l^{3}}}\int_{-\infty}^{\infty}dz\;z^{2}\,\Big(-G(z)\Big)\,\frac{\sin(2\pi z)}{2\pi z}\,e^{-z^{2}/(2\l)},
\ee
with $G(z)\equiv\sum_{m}\frac{\rho_{m}}{m}(4z^{2})^{m}$ where
\be
\la{B.4}
\rho_{m}\equiv r_{m}\zeta(2m) = \frac{4m}{B_{2m}}\cdot\frac{(-1)^{m+1}B_{2m}(2\pi)^{2m}}{2(2m)!} = \frac{2m(-1)^{m}(2\pi)^{2m}}{(2m)!}.
\ee
This gives 
\be
\la{B.5}
G(z) = 2\sum_{m\geq1}\frac{(-1)^{m}(4\pi z)^{2m}}{(2m)!} = 2\big[\cos(4\pi z)-1\big] = -4\sin^{2}(2\pi z).
\ee
Substituting \eqref{B.5} into \eqref{B.3}, using $\sin^{3}\theta=\frac14\big[3\sin\theta-\sin3\theta\big]$, gives
\be
\la{B.6}
-\Big(\sum_{m}\delta s_{m}\l^{m}\Big)e^{-2\pi^{2}\l} = \frac{1}{2\pi\sqrt{2\pi\l^{3}}}\int_{-\infty}^{\infty}dz\;z\big[3\sin(2\pi z)-\sin(6\pi z)\big]e^{-z^{2}/(2\l)}.
\ee
The integral $\int_{-\infty}^{\infty}z\sin(bz)\,e^{-z^{2}/(2\l)}dz=\sqrt{2\pi\l}\,\l b\,e^{-\l b^{2}/2}$ is obtained by differentiating
$\int_{-\infty}^{\infty} e^{ibz-z^{2}/2\l}dz=\sqrt{2\pi\l}\,e^{-\l b^{2}/2}$ with respect to $b$. We thus get 
\be
\la{B.7}
-\Big(\sum_{m}\delta s_{m}\l^{m}\Big)e^{-2\pi^{2}\l} = 3\big[e^{-2\pi^{2}\l}-e^{-18\pi^{2}\l}\big],
\ee
which is equivalent to \eqref{6.10}.

\subsection{Proof of $\delta s_m^{(2)}$}
\la{app:deltas2}

The same generating-function method used above to prove \eqref{6.10} extends to give a closed form for the second-order shift $\delta s_m^{(2)}$ defined in \eqref{9.2}, that is, the $O(\wt q^2)$ Taylor coefficient of
\be
\la{B.8}
\wh s_m(\wt q) \equiv s_m\big|_{\zeta_q(2k)\to\zeta(2k)E_{2k}(\wt q)},
\ee
the leading-$\l$ piece of $s_m(q)$ as a function of $\wt q$ alone. Since $\wh s_m(\wt q)$ is obtained from $s_m(q)$ by substituting each $\zeta_q(2k)$ with its own leading-$\l$ behavior, it is convenient to work directly with the generating function \eqref{6.2}, with $\zeta_q(2m)\to\zeta(2m)E_{2m}(\wt q)$ substituted throughout:
\ba
\la{B.9}
& \exp\Big(-\sum_m \wh s_m(\wt q)\,\l^m\Big) = \frac1{\sqrt{2\pi\l^3}}\int_{-\infty}^\infty dz\, z^2\, \wh F(z,\wt q)\, e^{-z^2/2\l}, \lp
\wh F(z,\wt q) \equiv \exp\Big(-\sum_m \frac{\zeta(2m) E_{2m}(\wt q)}m (4z^2)^m\Big).
\ea
Writing $E_{2m}(\wt q) = 1 + r_m\,\wt q + c_{2m,2}\,\wt q^2 + O(\wt q^3)$ for the Fourier expansion of the Eisenstein series, with $r_m$ the coefficient already used to define $\mc R$, we have
\be
\la{B.10}
\wh F(z,\wt q) = F^{\rm cl}(z)\,\exp\Big(-\wt q\, G(z) - \wt q^2 H(z) + O(\wt q^3)\Big), \qquad F^{\rm cl}(z) = \frac{\sin(2\pi z)}{2\pi z},
\ee
where $G(z) = -4\sin^2(2\pi z)$ is the function computed above, and
\be
\la{B.11}
H(z) \equiv \sum_{m\geq1} \frac{c_{2m,2}\,\zeta(2m)}m\,(4z^2)^m
\ee
is the analogous function built from the second Fourier coefficients. Using $\sigma_{2m-1}(2) = 1 + 2^{2m-1}$, we find that the explicit values are 
$c_{2m,2} = -\tfrac{4m}{B_{2m}}\,\sigma_{2m-1}(2) = r_m\,(1+2^{2m-1})$. This gives
\be
\la{B.12}
H(z) = \sum_m \frac{r_m\zeta(2m)}m(4z^2)^m + \frac12\sum_m \frac{r_m\zeta(2m)}m\big(4(2z)^2\big)^m = G(z) + \frac12\,G(2z),
\ee
and, by using the explicit form of $G(z)$, we get 
\be
\la{B.13}
H(z) = -4\sin^2(2\pi z) - 2\sin^2(4\pi z).
\ee
Expanding \eqref{B.10} to $O(\wt q^2)$, we get 
\be
\la{B.14}
\wh F(z,\wt q) = F^{\rm cl}(z)\Big[1 - \wt q\, G(z) + \wt q^2\Big(\frac{G(z)^2}2 - H(z)\Big) + O(\wt q^3)\Big],
\ee
so the $O(\wt q^2)$ piece of the integrand in \eqref{B.9} is $z^2 F^{\rm cl}(z)\big(\tfrac12 G(z)^2 - H(z)\big)e^{-z^2/2\l}$. Using now \eqref{B.13}, we obtain
\be
\la{B.15}
z^2 F^{\rm cl}(z)\Big(\frac{G(z)^2}2 - H(z)\Big) = \frac{48\,z}\pi \sin^3(\pi z)\cos^3(\pi z) = \frac z{2\pi}\Big[9\sin(2\pi z) - 3\sin(6\pi z)\Big],
\ee
where the last equality follows from $\sin\theta\cos\theta = \tfrac12\sin2\theta$ and $\sin^3\theta = \tfrac14(3\sin\theta-\sin3\theta)$. 
We now compute the Gaussian-sine moment term by term (see comment after (\ref{B.6})) and divide by $\sqrt{2\pi\l^3}$ to get
\be
\la{B.16}
\Big[\frac1{\sqrt{2\pi\l^3}}\int dz\, z^2\,\wh F(z,\wt q)\, e^{-z^2/2\l}\Big]_{O(\wt q^2)} = 9\,e^{-2\pi^2\l} - 9\,e^{-18\pi^2\l}.
\ee
Expanding the left-hand side of \eqref{B.9} to $O(\wt q^2)$ using $\wh s_m(\wt q) = s_m^{\rm cl} + \delta s_m\,\wt q + \delta s_m^{(2)}\,\wt q^2 + O(\wt q^3)$,
and recalling that only $s_1^{\rm cl}=2\pi^2$ survives among the classical values, the expression (\ref{B.16}) turns out to be equal to 
\be
\la{B.17}
e^{-2\pi^2\l}\bigg[\frac12\Big(\sum_m \delta s_m\,\l^m\Big)^2 - \sum_m \delta s_m^{(2)}\,\l^m\bigg].
\ee
Using, finally, $\sum_m\delta s_m\l^m = 3(e^{-16\pi^2\l}-1)$ from \eqref{6.10}, gives
\be
\sum_m \delta s_m^{(2)}\,\l^m = \frac12\Big[3\big(e^{-16\pi^2\l}-1\big)\Big]^2 - \Big[9 - 9\,e^{-16\pi^2\l}\Big] = \frac92\big(e^{-32\pi^2\l}-1\big),
\ee
which, expanded in $\l$, reproduces \eqref{9.3}:
\be
\delta s_m^{(2)} = \frac92\,\frac{(-32\pi^2)^m}{m!}.
\ee

\section{$\kappa$-class intersection numbers: the DVV recursion}
\la{app:DVV}

The $\kappa$-class intersection numbers entering \eqref{6.17} 
are computed by two standard, purely combinatorial recursions, which we record here in full, pedagogical detail.

\subsection{Pure $\psi$-intersection numbers} 

For $i=1,\dots,n$, $\psi_{i}\in H^{2}(\overline{\mc M}_{g,n};\mathbb Q)$ is the first Chern class of the line bundle whose fiber, 
over a point of moduli space, is the cotangent line to the curve at the $i$-th marked point. 
The symbol $\tau_{d}$ is standard shorthand (from the matrix-model/KdV literature in which these numbers 
first appeared \cite{Witten:1990hr,Kontsevich:1992ti}) for an insertion of $\psi^{d}$ at a marked point. 
Thus 
\be
\langle\tau_{d_{1}}\cdots\tau_{d_{n}}\rangle_{g}\equiv\int_{\overline{\mc M}_{g,n}}\psi_{1}^{d_{1}}\cdots\psi_{n}^{d_{n}},
\ee
which is nonzero only if $\sum_{i}d_{i}=3g-3+n$, matching the complex dimension of $\overline{\mc M}_{g,n}$.
Two base cases are
\be
\la{C.2}
\langle\tau_{0}^{3}\rangle_{0}=1,\qquad \langle\tau_{1}\rangle_{1}=\frac1{24},
\ee
together with the string and dilaton equations,
\be
\la{C.3}
\langle\tau_{0}\prod_{i}\tau_{d_{i}}\rangle_{g}=\sum_{j}\langle\tau_{d_{j}-1}\prod_{i\neq j}\tau_{d_{i}}\rangle_{g},
\qquad
\langle\tau_{1}\prod_{i}\tau_{d_{i}}\rangle_{g}=(2g-2+n)\langle\prod_{i}\tau_{d_{i}}\rangle_{g},
\ee
and the Witten--Kontsevich/DVV recursion \cite{Witten:1990hr,Kontsevich:1992ti,Dijkgraaf:1990rs} for the largest index $k_{1}\geq2$,
\ba
\la{C.4}
(2k_{1}{+}1)!!\,& \langle\tau_{k_{1}}\!\prod_{i=2}^{n}\tau_{d_{i}}\rangle_{g}
= \sum_{j=2}^{n}\frac{(2(k_{1}{+}d_{j}){-}1)!!}{(2d_{j}{-}1)!!}\langle\tau_{k_{1}+d_{j}-1}\prod_{i\neq1,j}\tau_{d_{i}}\rangle_{g}\lp
+\frac12 \sum_{a+b=k_{1}-2}(2a{+}1)!!(2b{+}1)!!\Big[\langle\tau_{a}\tau_{b}\prod_{i\geq2}\tau_{d_{i}}\rangle_{g-1} 
+\sum^{\rm stable}\langle\tau_{a}\prod_{I}\tau_{d_{i}}\rangle_{g_{1}}\langle\tau_{b}\prod_{J}\tau_{d_{i}}\rangle_{g_{2}}\Big].
\ea
The two terms in the second line of \eqref{C.4} correspond to the two ways a surface can degenerate, as described in 
Section~\ref{sec:bridge}: the non-separating pinch, which lowers the genus, and the separating one, which splits the 
surface into two stable pieces.
About notation, in \eqref{C.4}, $\sum^{\rm stable}$ denotes the sum over $g_{1}+g_{2}=g$ and over the splittings 
$I\sqcup J=\{2,\dots,n\}$, restricted to stable terms. Iterating \eqref{C.4} together with \eqref{C.2}, \eqref{C.3} 
generates every $\langle\tau_{d_{1}}\cdots\tau_{d_{n}}\rangle_{g}$.

Here $I\sqcup J=\{2,\dots,n\}$ means that the spectator marked points (all indices except $1$) are split into two 
\emph{disjoint} subsets $I$ and $J$ with $I\cup J=\{2,\dots,n\}$ and $I\cap J=\emptyset$, one subset going to each of 
the two resulting component curves -- the ``daughter surfaces'' produced when the original curve degenerates at a node 
into two separate pieces. The sum runs over all $2^{\,n-1}$ such splits (equivalently, over all subsets $I$, with $J$ 
the complement).
Each daughter surface also carries one further point -- $\tau_{a}$ or $\tau_{b}$ -- coming from the node where the two branches are glued back together.

Let us clarify what ``dropping unstable terms'' means here. We remind that $\overline{\mc M}_{g,n}$ is a well-defined moduli space only for $2g-2+n>0$; the four excluded cases are $(g,n)=(0,0),(0,1),(0,2),(1,0)$, 
where the underlying curve has a continuous automorphism group. Dropping unstable terms means that in the double sum we keep the term with $n_{1}=|I|+1$ points on the 
genus-$g_{1}$ side and $n_{2}=|J|+1$ on the genus-$g_{2}$ side only if
\be
\la{C.5}
2g_{1}-2+n_{1}>0 \qquad\text{and}\qquad 2g_{2}-2+n_{2}>0
\ee
both hold; otherwise that split contributes zero and is simply omitted from the sum. 
In practice, this is what makes the recursion terminate. For example, for $\langle\tau_{2}\tau_{0}\rangle_{1}$ there is only one spectator point to distribute, 
so every split puts $n_{1}\in\{1,2\}$ on the genus-$0$ side, giving $2(0)-2+n_{1}\leq0$ in every case, and the entire double-sum term vanishes identically.

\subsection{Mixed $\kappa$-class integrals} 

Using $\kappa_{a}=\pi_{*}(\psi_{n+1}^{a+1})$ and $\pi^{*}\kappa_{a}=\kappa_{a}-\psi_{n+1}^{a}$ under the forgetful map 
$\pi:\overline{\mc M}_{g,n+1}\to\overline{\mc M}_{g,n}$ \cite{arbarello1994}, one finds that removing one $\kappa_{a_{1}}$ from a product of $k$ requires expanding 
the pullback of the remaining $k-1$ factors over all subsets (not just pairwise merges), 
\be
\la{C.6}
\Big\langle\kappa_{a_{1}}\prod_{i=2}^{k}\kappa_{a_{i}}\Big\rangle_{g,n}
= \sum_{S\subseteq\{2,\dots,k\}}(-1)^{|S|}\Big\langle\tau_{a_{1}+1+\sum_{i\in S}a_{i}}\prod_{i\notin S}\kappa_{a_{i}}\Big\rangle_{g,n+1},
\ee
recursing on the (still $\kappa$-valued) product on the right until none remain, at which point the original $n$ points carry $\tau_{0}$ and \eqref{C.4} applies. 

Implemented in exact rational arithmetic, this reproduces every $\kappa$-number used anywhere here. 
Table~\ref{tab:kappa} collects, for every $(g,n)$ appearing in Tables~\ref{tab:322a}, \ref{tab:47a}, 
the complete set of top-degree pure $\kappa$-monomials, $\int_{\overline{\mc M}_{g,n}}\prod_m\kappa_m^{p_m}$ with $\sum_mm\,p_m=d=3g-3+n$ -- i.e.\ one entry per partition of $d$.

\begin{table}[htbp]
\centering
\def\arraystretch{1.3}
\begin{minipage}[t]{0.42\textwidth}
\vspace{0pt}
\centering
\begin{tabular}{cccc}
\toprule
$(g,n)$ & $d$ & $\kappa$-monomial & value \\
\midrule
$(0,4)$ & $1$ & $\kappa_1$ & $1$ \\
\midrule
$(1,1)$ & $1$ & $\kappa_1$ & $\frac1{24}$ \\
\midrule
$(0,5)$ & $2$ & $\kappa_1^2$ & $5$ \\
$(0,5)$ & $2$ & $\kappa_2$ & $1$ \\
\midrule
$(1,2)$ & $2$ & $\kappa_1^2$ & $\frac18$ \\
$(1,2)$ & $2$ & $\kappa_2$ & $\frac1{24}$ \\
\midrule
$(0,6)$ & $3$ & $\kappa_1^3$ & $61$ \\
$(0,6)$ & $3$ & $\kappa_1\kappa_2$ & $9$ \\
$(0,6)$ & $3$ & $\kappa_3$ & $1$ \\
\bottomrule
\end{tabular}
\end{minipage}%
\begin{minipage}[t]{0.58\textwidth}
\vspace{0pt}
\centering
\begin{tabular}{cccc}
\toprule
$(g,n)$ & $d$ & $\kappa$-monomial & value \\
\midrule
$(2,1)$ & $4$ & $\kappa_1^4$ & $\frac{29}{128}$ \\
$(2,1)$ & $4$ & $\kappa_1^2\kappa_2$ & $\frac{259}{5760}$ \\
$(2,1)$ & $4$ & $\kappa_1\kappa_3$ & $\frac{13}{1920}$ \\
$(2,1)$ & $4$ & $\kappa_2^2$ & $\frac{53}{5760}$ \\
$(2,1)$ & $4$ & $\kappa_4$ & $\frac1{1152}$ \\
\midrule
$(2,2)$ & $5$ & $\kappa_1^5$ & $\frac{787}{128}$ \\
$(2,2)$ & $5$ & $\kappa_1^3\kappa_2$ & $\frac{1073}{1152}$ \\
$(2,2)$ & $5$ & $\kappa_1^2\kappa_3$ & $\frac{41}{384}$ \\
$(2,2)$ & $5$ & $\kappa_1\kappa_2^2$ & $\frac{827}{5760}$ \\
$(2,2)$ & $5$ & $\kappa_1\kappa_4$ & $\frac{59}{5760}$ \\
$(2,2)$ & $5$ & $\kappa_2\kappa_3$ & $\frac{97}{5760}$ \\
$(2,2)$ & $5$ & $\kappa_5$ & $\frac1{1152}$ \\
\bottomrule
\end{tabular}
\end{minipage}
\caption{Every top-degree pure $\kappa$-monomial, $\int_{\overline{\mc M}_{g,n}}\prod_m\kappa_m^{p_m}$ with 
$\sum_m m\,p_m=d$, one per partition of $d$, for the $(g,n)$ appearing in Tables~\ref{tab:322a}--\ref{tab:NP2}: 
the five cases with $d\leq3$ on the left, $(2,1)$ and $(2,2)$ on the right. Computed via \eqref{C.4}, \eqref{C.6} 
and cross-checked with the \texttt{admcycles} package in SageMath \cite{Delecroix2021}; the $(2,1)$ block agrees 
with \eqref{6.13}.}
\la{tab:kappa}
\end{table}

\section{Constant terms at $(0,6)$ and $(2,2)$}
\la{app:discrepancy}

For $(g,n)=(0,6)$ and $(2,2)$, the expressions for $V^q_{g,n}(0)$ obtained from \eqref{6.1} are not the same as those 
tabulated in v1 of \cite{DoNorbury:2025dssyk}. We are grateful to N.~Do and P.~Norbury for correspondence confirming 
that those expressions contain typographical errors and that the ones recorded below are the correct ones. This appendix collects 
them.

It is worth noting why the difference is not visible classically. Since $s_1(q)\to2\pi^2$ and $s_m(q)\to0$ for 
$m\geq2$ as $q\to1$, the constant term $V^q_{g,n}(0)$ reduces in that limit to a multiple of 
$\int_{\overline{\mc M}_{g,n}}\kappa_1^{d}$ alone, and is insensitive to the coefficients of every $\kappa_m$ with 
$m\geq2$. The leading non-perturbative coefficient, by contrast, involves the full set $\int\kappa_m\kappa_1^{d-m}$ 
through \eqref{6.17}. The two cases below illustrate this: in each, both expressions reproduce the classical 
Weil--Petersson volume exactly, and they are distinguished only by $\sssb_{g,n}$.

\paragraph{The case $(0,6)$.} Here $d=3$ and the three $\kappa$-numbers are those of Table~\ref{tab:kappa}. Assembling,
\be
\la{D.1}
V^q_{0,6}(0)=\frac{s_1^3}6\cdot61+s_1s_2\cdot 9+s_3\cdot 1
= 12640\,\zeta_q(2)^3+11040\,\zeta_q(2)\zeta_q(4)+2240\,\zeta_q(6),
\ee
against the tabulated $17824\,\zeta_q(2)^3-1920\,\zeta_q(2)\zeta_q(4)+2240\,\zeta_q(6)$. The two agree in the 
$\zeta_q(6)$ coefficient and differ by
\be
\la{D.2}
5184\Big[\zeta_q(2)^3-\tfrac52\,\zeta_q(2)\zeta_q(4)\Big],
\ee
which vanishes at $\zeta_q(2k)\to\zeta(2k)$, since $\zeta(2)^3=\tfrac52\zeta(2)\zeta(4)=\pi^6/216$. Both expressions 
therefore give the correct classical volume $V^{\rm WP}_{0,6}(0)=\tfrac{244}3\pi^6$, and differ only in the 
non-perturbative sector: \eqref{D.1} gives $\sssb_{0,6}=-992$, the tabulated expression $-7904$.

\paragraph{The case $(2,2)$.} Here $d=5$, so seven monomials occur, one per partition of $5$; the tabulated expression 
contains five of them, with no $\zeta_q(2)\zeta_q(8)$ or $\zeta_q(10)$ term. The seven $\kappa$-numbers are those 
of Table~\ref{tab:kappa}. Assembling at degree $5$ requires $s_5(q)$, hence $\zeta_q(10)$, for which \eqref{3.14}, 
\eqref{3.15} give
\be
\la{D.3}
\zeta_{q}(10) = \frac{14797-12672E_{2}-1804E_{4}-286E_{6}-33E_{4}^{2}-2E_{4}E_{6}}{191600640},
\ee
and \eqref{6.2} gives
\ba
\la{D.4}
s_{5}(q) &= \tfrac{49152}{5}\zeta_{q}(2)^{5}-245760\,\zeta_{q}(2)^{3}\zeta_{q}(4)
+860160\,\zeta_{q}(2)^{2}\zeta_{q}(6)+737280\,\zeta_{q}(2)\zeta_{q}(4)^{2}\lp
-1935360\,\zeta_{q}(2)\zeta_{q}(8)-1505280\,\zeta_{q}(4)\zeta_{q}(6)+2128896\,\zeta_{q}(10),
\ea
which vanishes at the classical values, as required. The result is
\ba
\la{D.5}
V^{q}_{2,2}(0) &= 3439\,\zeta_{q}(2)^{5}+12194\,\zeta_{q}(2)^{3}\zeta_{q}(4)+6333\,\zeta_{q}(2)\zeta_{q}(4)^{2}
+9548\,\zeta_{q}(2)^{2}\zeta_{q}(6)\lp
+3220\,\zeta_{q}(4)\zeta_{q}(6)+5754\,\zeta_{q}(2)\zeta_{q}(8)+1848\,\zeta_{q}(10).
\ea
As at $(0,6)$, the difference from the tabulated expression vanishes at $\zeta_q(2k)\to\zeta(2k)$: both give 
$V^{\rm WP}_{2,2}(0)=\tfrac{787}{480}\pi^{10}$. In the non-perturbative sector \eqref{D.5} gives 
$\sssb_{2,2}=-\tfrac{2183}{180}$, the tabulated expression $-\tfrac{7609}{12}$.

\bibliography{Krylov-Biblio}
\bibliographystyle{JHEP}
\end{document}